\documentclass[aps,twocolumn,showpacs,amsmath,amssymb,superscriptaddress,prc]{revtex4-2}
\usepackage{graphicx,here}
\usepackage{multirow}
\usepackage{tikz}
\usepackage{epstopdf}
\usepackage[percent]{overpic}
\usepackage{scrextend}
\usepackage{xcolor,xspace}
\usepackage[ulem=normalem]{changes}
\usepackage{hyperref}
\hypersetup{colorlinks=True,urlcolor=blue,linkcolor=blue,citecolor=blue,filecolor=black}

\colorlet{Changes@Color}{red}
\usepackage{dcolumn,color,footnote,bm,braket}
\usepackage{url,longtable,tabularx}
\usepackage{threeparttable}

\usepackage{fancybox}
\usetikzlibrary{matrix}

\begin{document}

\title{Impacts of hexadecapole correlations in actinide nuclei}

\author{L. Lotina}
\email{llotina@imi.hr}
\affiliation{Division of Radiation Protection, Institute for Medical Research and Occupational Health, HR-10000 Zagreb, Croatia}

\author{K. Nomura}
\email{nomura@sci.hokudai.ac.jp}
\affiliation{Department of Physics, 
Hokkaido University, Sapporo 060-0810, Japan}
\affiliation{Nuclear Reaction Data Center, 
Hokkaido University, Sapporo 060-0810, Japan}

\author{R. Rodr\'{\i}guez-Guzm\'an}
\email{guzman.rodriguez@nu.edu.kz}
\affiliation{Department of Physics, School of Sciencies and Humanities, Nazarbayev 
University, 53 Kabanbay Batyr Ave., Astana 010000, Kazakhstan}

\author{L. M. Robledo}
\affiliation{Departamento de F\'{i}sica Te\'{o}rica and CIAFF, Universidad Aut\'{o}noma de Madrid, E-28049 Madrid, Spain}
\affiliation{Center for Computational Simulation, Universidad Polit\'{e}cnica de Madrid, Campus de Montegancedo, Bohadilla del Monte, E-28660-Madrid, Spain}

\date{\today}

\begin{abstract}
The impact of hexadecapole correlations on the low-energy
spectroscopic properties of Th, U, and Pu nuclei, within the mass range 
$232 \le A \le 240$, is studied systematically using the mapped $sdg$-IBM
model. Fermionic input is obtained via the  quadrupole-hexadecapole constrained
Hartree-Fock-Bogoliubov approximation, based on the parametrization D1S of the 
Gogny energy density functional. The $sdg$-IBM Hamiltonian parameters are
determined  by mapping
the quadrupole-hexadecapole fermionic mean-field potential energy
surfaces 
onto the corresponding bosonic surfaces. The  low-energy spectra
and transition strengths, obtained via the diagonalization of the $sdg$-IBM
Hamiltonian, compare well with the available experimental data. It is shown
that the effects of  hexadecapole collectivity
can be observed in high-spin yrast states
with spins $J^{\pi} \geqslant 10^{+}$. The mapped $sdg$-IBM improves 
the excitation energies of those states, as compared with the 
simpler $sd$-IBM model. The $sdg$-IBM also improves the 
description of the $E2$ transition strengths
between high-spin yrast states and 
predicts strong $E4$ transitions
from nonyrast $4^+$ states to
the $0^+$ ground state.
\end{abstract}

\maketitle

\section{Introduction}

Intrinsic deformations represent a prominent aspect of 
atomic nuclei with an impact on several nuclear structure
properties \cite{BM}. Within this context, the most common 
intrinsic deformation is of quadrupole type. Such quadrupole
deformations have been extensively studied and their 
fingerprints on the low-energy
positive-parity states are well known. On the other hand, the 
(second-order) hexadecapole deformations also contribute to
the positive-parity states \cite{BM}. However, as both 
the quadrupole and hexadecapole deformations affect 
states with the same parity quantum numbers, the 
dominant quadrupole correlations
often overshadow the hexadecapole correlation effects.
Nevertheless, hexadecapole correlations
have already been observed in different regions 
of the nuclear chart such as the light 
\cite{gupta2020}, medium-heavy \cite{spieker2023},
rare-earth \cite{erb1972, wollersheim1977},
transitional metals
\cite{baker1989, sethi1991},
and actinide \cite{bemis1973,zumbro1984} regions. 
The $K^{\pi}=4^+$
bands at low energy, which exhibit pronounced
$E4$ transitions to the ground-state band, provide 
characteristic observables 
associated with hexadecapole correlations.
Recently, a significant interest has
been devoted to studying the hexadecapole
deformation effects in
relativistic heavy ion collisions
of actinide nuclei,
especially in $^{238}$U
\cite{ryssens2023, xu2024, schenke2024}.
It is therefore timely and relevant
to conduct a systematic theoretical
study of hexadecapole deformation effects
in actinide nuclei using a microscopic model.

One of the most commonly employed
models to describe
collective properties in nuclei is
the interacting boson model (IBM)
\cite{IBM}. In the simplest version
of the IBM, called the $sd$-IBM,
a nucleus is approximated as a system
composed of an inert core, defined by the nearest double-magic nucleus,
and valence nucleons that couple
into the correlated monopole
(with spin and parity $0^+$)
and quadrupole ($2^+$) pairs,
which are represented by
$s$ and $d$ bosons, respectively
\cite{OAIT,OAI}.
The interactions between those
bosons give rise to the low-energy quadrupole
collective states.
The hexadecapole degree of freedom
is introduced in the IBM via the inclusion of
$g$ bosons ($4^+$), corresponding to
the hexadecapole pairs of valence nucleons.
This extension is called the $sdg$-IBM,
and it has already been used in previous studies
\cite{casten1988,otsuka1981,otsuka1982,otsuka1985,otsuka1988, devi-kota1990, kuyucak1994, vanisacker2010}.

Originally, the IBM was built as a phenomenological model and the 
parameters of the IBM Hamiltonian had to be fitted to experimental data 
corresponding to the low-lying energy spectrum for each nucleus. In the 
last decades a fermion-to-boson mapping procedure has 
been developed
\cite{nomura2008,nomura2025rev}
that allows  to derive the
parameters of the IBM Hamiltonian microscopically, i.e., by fitting them to the potential energy surface (PES)
obtained within the constrained self-consistent
mean-field (SCMF) approximation, based on a given universal
energy density functional (EDF)
\cite{bender2003,vretenar2005,robledo2019}.
The fermion-to-boson mapping procedure has been successfully
employed to describe the effects of
axial and triaxial quadrupole \cite{nomura2008, nomura2010, nomura2011pt, nomura2012tri}
as well as octupole \cite{nomura2013oct, nomura2014,nomura2023oct} correlations 
in atomic nuclei. The mapped IBM has also been employed in the analysis 
of hexadecapole correlations
in rare-earth nuclei
\cite{lotina2024hex-1,lotina2024hex-2,lotina2025}. For a detailed account 
of the fermion-to-boson mapping procedure the reader is referred to 
Ref.~\cite{nomura2025rev} and references therein.

In this work, we extend the analyses
of the quadrupole-hexadecapole
mapping procedure, already carried out for 
rare-earth nuclei \cite{lotina2024hex-1,lotina2024hex-2,lotina2025}, to 
the actinide region of the nuclear chart. To this end, we have considered the isotopic 
chains $^{232-240}$Th $(Z=90)$, $^{232-240}$U $(Z=92)$,
and $^{232-240}$Pu $(Z=94)$. As an input, we have considered the 
Hartree-Fock-Bogoliubov (HFB)
approximation, based on the Gogny-D1S \cite{GOGNY1975,BERGER1984}
EDF. This microscopic mean field framework has been extensively
applied to nuclei
across the whole  nuclear chart
\cite{hilaire2007,robledo2019}. It has also been successfully combined
with the IBM mapping procedure already  mentioned
above to analyze the impact of the quadrupole
\cite{nomura2011pt,nomura2011Os,nomura2011sys,
nomura2017ge,nomura2017kr},
and octupole
\cite{nomura2015,nomura2020oct,nomura2021oct-u,
nomura2021oct-ba,nomura2021oct-zn}
correlations on the spectroscopic properties of
even-even nuclei.

The quadrupole-hexadecapole coupling has been studied 
recently at the HFB level with the help of
constrains on the (axial) quadrupole and hexadecapole 
operators in Sm, Gd, rare-earth and actinide nuclei
\cite{kumar-robledo2023,
guzman2025,guzman2025rare_earth}. Those previous studies 
\cite{kumar-robledo2023,
guzman2025,guzman2025rare_earth}
have also considered the role of the  
zero-point 
quantum quadrupole-hexadecapole fluctuations within 
the framework of the two-dimensional (2D) Generator 
Coordinate Method (GCM) 
\cite{RS,bender2003,niksic2011,robledo2019}. Both at the 
HFB and 2D-GCM 
levels, calculations have been carried out in terms 
of the Gogny-D1S EDF. In this paper, we will consider 
an alternative and more detailed description of spectroscopic properties 
along the selected Th, U and Pu isotopic chains. In 
particular, we examine the role of the
hexadecapole
degree of freedom in the low-lying states of 
the selected set of even-even Th, U and Pu 
nuclei within the framework of 
the Gogny-D1S HFB mapped $sdg$-IBM model. To better undertand the 
role of hexadecapole correlations, comparisons will be 
presented with results obtained within the 
Gogny-D1S HFB mapped $sd$-IBM model.

It is worth mentioning that the
considered nuclei 
have been shown to exhibit
some degree of octupole collectivity
\cite{butler1996,butler2016} which in turn 
might have certain influence 
on the positive-parity
states, such as the excited $0^+$
states
\cite{spieker2013,nomura2014,spieker2018,nomura2020oct}. In this
study, we neglect such octupole correlation effects not only 
for the sake of simplicity, but also because the octupole degree of freedom
is more relevant for the properties
of  negative-parity states.

The paper is organized as follows. The theoretical 
framework used is briefly outlined 
in Sec. \ref{sec:model}. Such a theoretical scheme comprises 
the Gogny-D1S SCMF calculations as well as the 
mapped $sdg$-IBM and $sd$-IBM.
The (axial) quadrupole-hexadecapole
PESs obtained from the SCMF
and IBM calculations are discussed in Sec.~\ref{sec:pes}. The 
results of the spectroscopic calculations
are presented in Sec.~\ref{sec:results}.
Finally, Sec.~\ref{sec:summary} is devoted to the 
concluding remarks and work perspectives.

%
\section{Model description\label{sec:model}}

For the selected set of Th, U and Pu nuclei, the 
SCMF PESs are obtained within the constrained 
Gogny-D1S HFB approximation \cite{GOGNY1975,BERGER1984}. 
Constrains on the (axial) quadrupole $\hat{Q}_{20}$
and hexadecapole $\hat{Q}_{40}$ operators are used to explore the PES for the
two degrees of freedom. It is convenient to use mass independent deformation 
parameters $\beta_{\lambda 0}$ ($\lambda=2,4$) which are defined in the standard way as 
\cite{kumar-robledo2023,
guzman2025,guzman2025rare_earth}
\begin{equation}
\beta_{\lambda0}
=\frac{\sqrt{4 \pi (2 \lambda + 1)}}{3 R^{\lambda} A}
\braket{\hat{Q}_{\lambda 0}}
\; ,
    \label{eq1}
\end{equation}
where  
$R=1.2 A^{1/3}$ fm is the radius and $A$
is the mass number. For simplicity, we denote $\beta_{\lambda 0}$
as $\beta_{\lambda}$, since  axial symmetry
is kept as a self-consistent symmetry in the Gogny-HFB calculations. In 
Eq.(\ref{eq1}), $\braket{\hat{Q}_{\lambda 0}}$ represents 
the average value  of quadrupole and/or hexadecapole multipole operator
in the corresponding HFB state. More 
details on the $(\beta_2,\beta_4)$-constrained
Gogny-HFB calculations can be found in Ref.~\cite{guzman2025}.

In order to obtain low-energy spectra
and transition properties, calculations have been carried out 
within the (mapped) $sdg$-IBM. For the $sdg$-IBM model
Hamiltonian, we have assumed the same form as in previous studies
\cite{lotina2024hex-1,lotina2024hex-2,lotina2025}:
\begin{equation}
    \hat{H} = \epsilon_d \hat{n}_d + \epsilon_g \hat{n}_g + \kappa \hat{Q}^{(2)} \cdot \hat{Q}^{(2)} + \kappa(1- \chi^2) \hat{Q}^{(4)} \cdot \hat{Q}^{(4)}.
    \label{eq:hb_sdg}
\end{equation}
where
$\hat{n}_d= d^{\dagger} \cdot \tilde{d}$ and $\hat{n}_g= g^{\dagger} \cdot \tilde{g}$ represent the $d$- and $g$-boson number operators, respectively, while
\begin{equation}
\begin{aligned}
  \hat{Q}^{(2)} = &(s^{\dagger} \times \tilde{d} + d^{\dagger} \times s) + \chi \left[\frac{11 \sqrt{10}}{28}(d^{\dagger} \times \tilde{d})^{(2)}\right. \\
      &
\left.
-\frac{9}{7} \sigma (d^{\dagger} \times \tilde{g} + g^{\dagger} \times \tilde{g})^{(2)}
+ \frac{3 \sqrt{55}}{14} (g^{\dagger} \times \tilde{g})^{(2)} \right], 
\end{aligned}
\label{eq3}
\end{equation}
and
\begin{equation}
    \hat{Q}^{(4)} =  s^{\dagger} \times \tilde{g} + g^{\dagger} \times s
    \label{eq4}
\end{equation}
stand for the boson image of the quadrupole and hexadecapole
operators, respectively. In principle, the $sdg$-boson model
Hamiltonian should include more interaction terms
and the corresponding parameters \cite{IBM}. However, as already shown 
in previous studies \cite{lotina2024hex-1,lotina2024hex-2,lotina2025}, the 
Hamiltonian Eq.(\ref{eq:hb_sdg}) provides 
a reasonable starting point to describe the properties of nuclei
for which the experimental data suggest a pronounced 
quadrupole-hexadecapole coupling.
It also satisfies the
$\textnormal{U(5)}\otimes\textnormal{U(9)}$,
SU(3) and O(15) dynamical symmetries
\cite{kota1987,vanisacker2010}.
For $\sigma=1$, the Hamiltonian Eq.(\ref{eq:hb_sdg})
is equivalent to the one considered
in Ref.~\cite{vanisacker2010}. In order to guarantee
the physical meaning of each term of the $sdg$-IBM 
Hamiltonian, it is assumed that the quadrupole 
interaction is attractive ($\kappa<0$) and that 
the energy required
for two valence nucleons to couple into a $d$-boson is smaller than 
the energy required for them to couple into a $g$-boson
($\epsilon_d < \epsilon_g$). Furthermore, as the considered actinide 
nuclei display strong axial deformations, they are assumed to be 
close to the
SU(3) limit of the $sdg$-IBM. To meet the requirement for the Hamiltonian to be in the SU(3)
symmetry limit, the parameters
$\lambda$ and $\sigma$ must satisfy
the conditions
$-1 \leqslant \chi \leqslant +1$
and $-1 \leqslant \chi \sigma \leqslant +1$
\cite{vanisacker2010}.

The parameters $\epsilon_d$, $\epsilon_g$,
$\kappa$, $\chi$ and $\sigma$ are
determined using the $sdg$-IBM mapping
procedure developed in Ref.~\cite{lotina2024hex-1}.
To connect the $sdg$-IBM to the geometric
model of the nucleus \cite{ginocchio1980,dieperink1980},
a coherent state (up to the normalization factor)
\begin{equation}
\label{eq:coherent}
 \ket{\phi} = (1+ \tilde{\beta}_2 d^{\dagger}_0 + \tilde{\beta}_4 g^{\dagger}_0)^{N_B} \ket{0}
\; ,
\end{equation}
is introduced, where $N_B$  and $\ket{0}$ represent 
the number of bosons and the boson vacuum (or the inert core). 
On the other hand, $\tilde{\beta}_2$ and $\tilde{\beta}_4$
are the axial bosonic quadrupole 
and hexadecapole deformations, which, as shown below, are related to the
fermionic deformations of Eq.~(\ref{eq1}). 
For all the considered  isotopes, the inert core corresponds to the doubly-magic nucleus $^{208}_{82}$Pb,
while the number of bosons
is equal to $N_B=12-16$ for the
even-even nuclei with $A=232-240$.
The $sdg$-IBM PES, as a function of the 
$\tilde{\beta}_2$ and $\tilde{\beta}_4$ deformations, 
is obtained
as the expectation value of the
Hamiltonian \eqref{eq:hb_sdg}
in the coherent state $\braket{\phi}$
\eqref{eq:coherent}, i.e.,
$E_{\textnormal{IBM}}(\tilde{\beta}_2,\tilde{\beta}_4)=\bra{\phi} \hat{H} \ket{\phi}/\braket{\phi|\phi}$.
The exact analytical form of the $sdg$-IBM PES
can be found in
Refs.~\cite{vanisacker2010, devi-kota1990}.

The parameters of the
$sdg$-IBM are determined
so that the bosonic PES,
$E_{\textnormal{IBM}}(\tilde{\beta}_2,\tilde{\beta}_4)$,
matches as close as possible the Gogny-D1S HFB PES,
$E_{\textnormal{SCMF}}({\beta}_2,{\beta}_4)$,
near
the absolute minimum:
\begin{equation}
    E_{\textnormal{SCMF}}(\beta_2, \beta_4) \approx E_{\textnormal{IBM}}(\tilde{\beta}_2, \tilde{\beta}_4) \; .
    \label{eq5}
\end{equation}
The standard assumption of the
IBM mapping procedure is that
the bosonic axial deformations should be
proportional to those of the SCMF calculations, i.e.,
\begin{equation}
\label{eq:cbeta}
\tilde{\beta}_2 = C_2 \beta_2\,
\quad ,
\quad
\tilde{\beta}_4 = C_4 \beta_4 
\; ,
\end{equation}
with $C_{\lambda}$ being the proportionality
constants.
These relations have already been shown
to work well in previous calculations
including (axial)  octupole
\cite{nomura2013oct,nomura2014}
and hexadecapole
\cite{lotina2024hex-1,lotina2024hex-2}
deformations.
At the end, one should determine seven parameters,
5 in the $sdg$-IBM Hamiltonian,
and the two scale factors, $C_2$ and $C_4$.

An important point to mention regarding the 2D mapping method with axial deformation parameters as coordinates is that the obtained PESs from the IBM mapping tend to be flat compared to their SCMF counterparts. This is observed both in the case of quadrupole-octupole \cite{nomura2013oct, nomura2014} and quadrupole hexadecapole mapping \cite{lotina2024hex-1,lotina2024hex-2}. This is to be expected, as IBM has a significantly smaller model space than the SCMF models. For the mapping method to be considered successful, certain conditions must be met: first, the positions of the absolute minima of the SCMF PES must be reproduced. Furthermore, IBM PES should also reproduce the overall shape of the SCMF PES as best as possible, and it should also reproduce the existence of low-energy saddle points in the SCMF PES, if there are any. Thirdly, the IBM PES needs to be as close in size as possible to the SCMF counterpart. Finally, all of the parameters must have values consistent with the physical conditions and symmetry limits imposed by the IBM Hamiltonian. Since small variations in the IBM parameters can lead to significant changes in the locations of absolute minima, PES sizes, and saddle point positions, the obtained sets of IBM parameters for each nucleus can be considered to be approximately unique, with possible variations in parameter values being sufficiently small so as not to affect any of the spectroscopic results significantly.

For the $sd$-IBM calculations, the 
following simple Hamiltonian is employed \cite{IBM}:
\begin{equation}
    \hat{H}_{sd} = \epsilon_d \hat{n}_d + \kappa \hat{Q}^{(2)} \cdot \hat{Q}^{(2)},
    \label{eq:hb_sd}
\end{equation}
with the boson quadrupole operator
$\hat Q^{(2)}$ being given by
\begin{equation}
    \hat{Q}^{(2)} = s^{\dagger} \tilde{d} + d^{\dagger} s +
\chi ( d^{\dagger} \times \tilde{d})^{(2)}.
    \label{eq7}
\end{equation}
The parameters $\epsilon_d$, $\kappa$, and $\chi$ 
are fitted so that a one-dimensional (1D),
$sd$-IBM energy curve should approximately
reproduce the Gogny-D1S PES
along the $\beta_4=0$ line:
\begin{equation}
E_{\textnormal{SCMF}}(\beta_2, \beta_4=0)
\approx
E_{sd-\textnormal{IBM}}(\tilde{\beta}_2)
\; .
\label{eq8}
\end{equation}
Within  the $sd$-IBM mapping,
attention is paid to reproduce
the positions of the absolute minima,
the depth of the potential,
and the energy difference between the
minimum on the $\beta_4=0$ axis and the saddle point
on the oblate side of the HFB PES.
A similar relationship between
the fermionic and bosonic deformations
to Eq.~\eqref{eq:cbeta}
is assumed for the $sd$-IBM, that is,
$\tilde{\beta}_2 = C^{sd}_2 \beta_2$,
leaving us with a total of 4 parameters
to be determined.

Low-energy spectra and
transition strengths have been
obtained by diagonalizing
the mapped boson Hamiltonians
Eqs. \eqref{eq:hb_sdg} and \eqref{eq8}.
The numerical diagonalization is performed
using the computer program ARBMODEL
\cite{arbmodel}.

The bosonic quadrupole $\hat T(E2)$
and hexadecapole $\hat T(E4)$
transition operators are
formulated for both the $sdg$-IBM
and $sd$-IBM in the same way as in
Refs.~\cite{lotina2024hex-1,lotina2024hex-2,lotina2025}.
The bosonic quadrupole transition operator reads
\begin{equation}
\hat{T}(E2)_{sdg/sd} = e_2^{sdg/sd} \hat{Q}^{(2)}_{sdg/sd},
\label{eq9}
\end{equation}
with $\hat{Q}^{(2)}_{sdg}$
and $\hat{Q}^{(2)}_{sd}$ being the
quadrupole operators of the $sdg$-IBM
Eq.~(\ref{eq3}) and $sd$-IBM
Eq.~(\ref{eq7}), with the same values
of the parameters $\chi$ and $\sigma$
as those used for these operators.
In the $sdg$-IBM, $\hat T(E4)_{sdg}$
is given as
\begin{equation}
    \hat{T}^{(E4)}_{sdg} = e_4^{sdg} \left [  s^{\dagger} \tilde{g} + g^{\dagger} s + (d^{\dagger} \times \tilde{d})^{(4)} \right ],
    \label{eq10}
\end{equation}
while, for the $sd$-IBM,
it is of the form:
\begin{equation}
    \hat{T}^{(E4)}_{sd} = e_4^{sd} (d^{\dagger} \times \tilde{d})^{(4)}.
    \label{eq11}
\end{equation}
The quadrupole effective charge $e_2^{sdg/sd}$
is determined for each nucleus by fitting
the calculated quadrupole transition
strength, $B(E2; 2_1^+ \rightarrow 0_1^+)$,
to the experimental data.
Similarly, the hexadecapole effective charge
$e_4^{sdg/sd}$ is fitted so that the
experimental data on the
$B(E4; 4_1^+ \rightarrow 0_1^+)$
transition strength in the yrast band is reproduced.
Since data on $E4$ transitions
in the actinide region are limited,
this calculation is only performed
for nuclei for which experimental
$B(E4)$ values are available.

\begin{figure*}    
\begin{center}
\includegraphics[width =\textwidth]{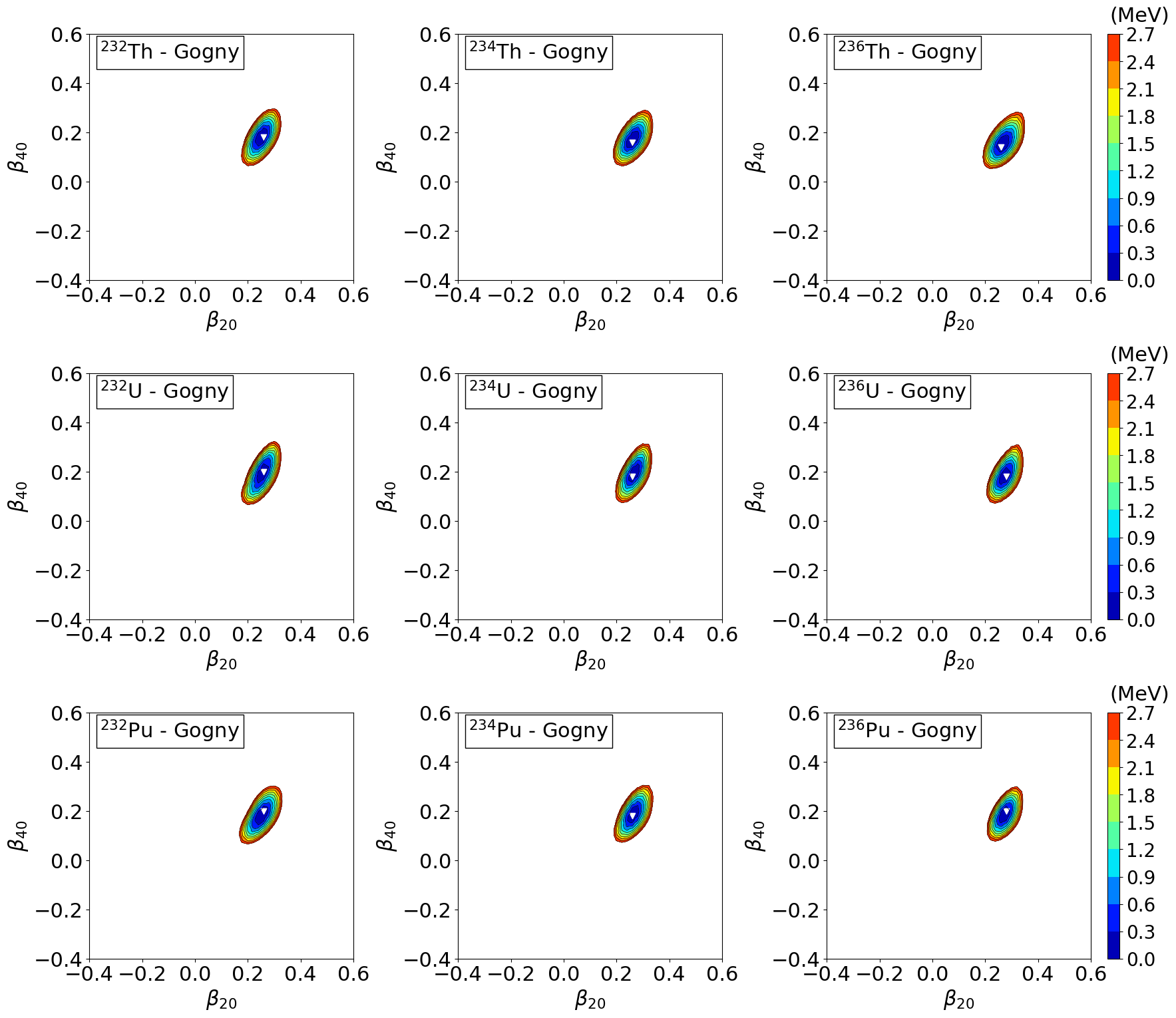}
\caption{(Color online) The Gogny-D1S PESs obtained for the nuclei
$^{232-236}$Th (first row),
$^{232-236}$U (second row) and  $^{232-236}$Pu (third row)
are plotted as functions of the quadrupole and hexadecapole 
deformations. The energy difference between neighboring contours is
0.1 MeV.
The absolute minima are indicated by
open triangles.} 
\label{GOGNY_PES}
\end{center}
\end{figure*}


\begin{figure*}    
\begin{center}
\includegraphics[width = .7\linewidth]{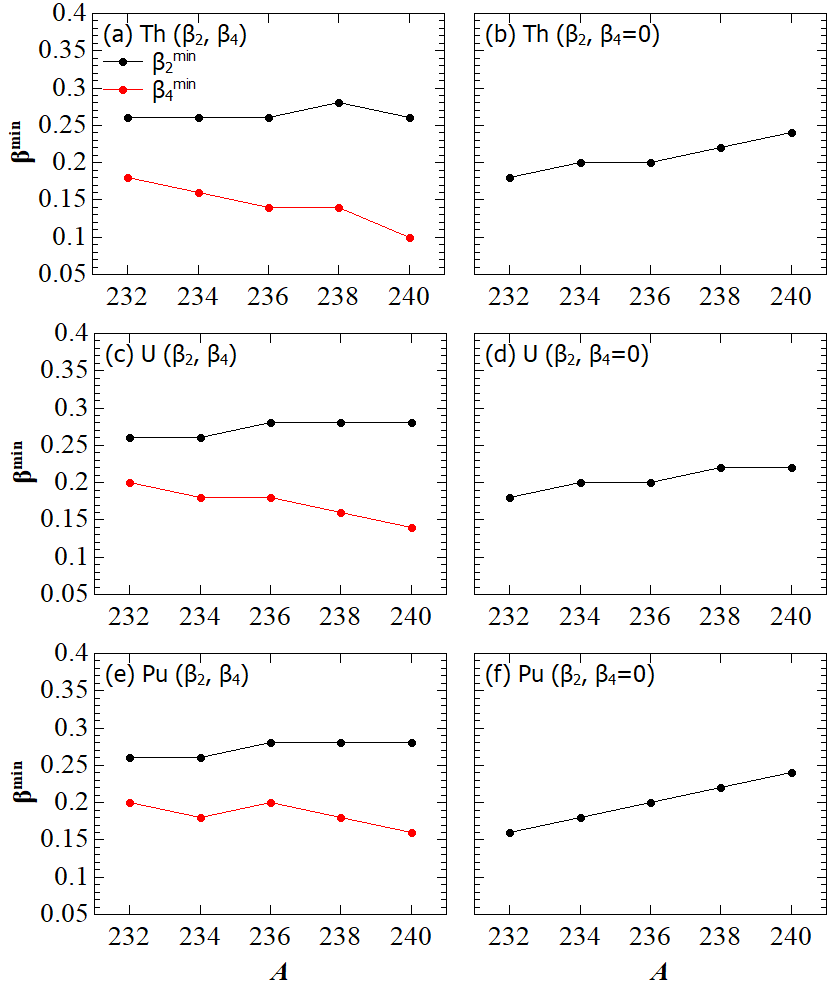}
\caption{(Color online) The quadrupole $\beta_2^{\textnormal{min}}$ and 
hexadecapole $\beta_4^{\textnormal{min}}$ deformations 
corresponding to the absolute minima of the Gogny-D1S HFB PESs
in Fig.~\ref{GOGNY_PES} are plotted as functions of the mass number 
$A$ for the nuclei $^{232-240}$Th [panel (a)],
$^{232-240}$U [panel (c)] and  $^{232-240}$Pu [panel (e)]. The 
$\beta_2^{\textnormal{min}}$ deformations 
corresponding to the absolute minima of the PESs in 
Fig.~\ref{GOGNY_PES} along the $\beta_4=0$ line
are depicted in panels (b), (d) and (f).
} 
\label{fig:beta_min}
\end{center}
\end{figure*}

\begin{figure*}
\begin{center}
\includegraphics[width =\textwidth]{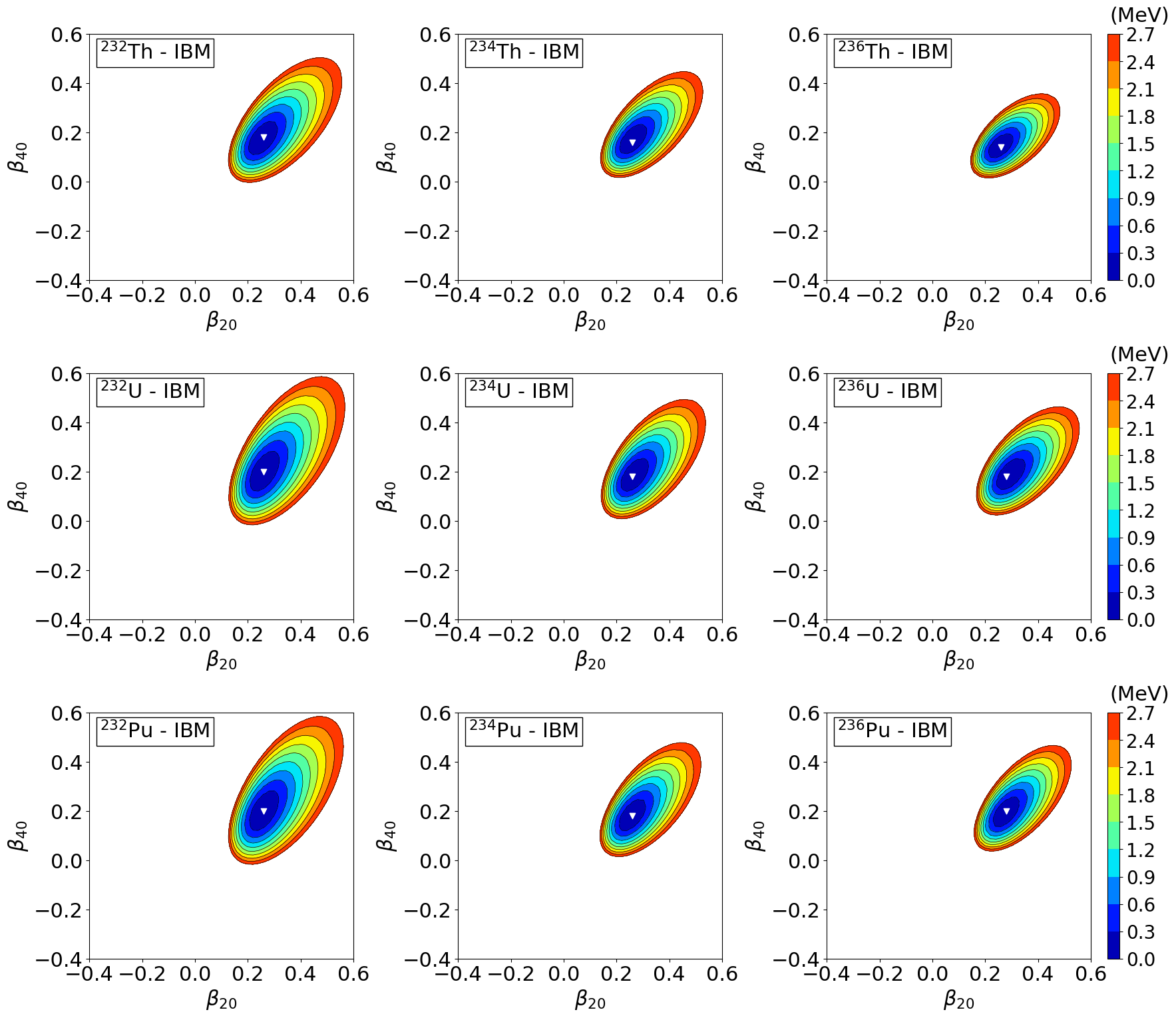}
\caption{(Color online) The same as in  Fig. \ref{GOGNY_PES}, but for the mapped $sdg$-IBM PESs of $^{232-236}$Th (first row),
$^{232-236}$U (second row) and  $^{232-236}$Pu (third row).} 
\label{IBM_PES}
\end{center}
\end{figure*}

\section{Mapping the SCMF results onto the IBM space\label{sec:pes}}

The Gogny-D1S PESs  obtained for even-even Th, U and Pu isotopes 
with mass number $A=232-236$ are depicted in Fig.~\ref{GOGNY_PES}.
Those PESs are shown up to an energy of 2.7 MeV above
the HFB ground state. Note, that the PESs for nuclei
with $A=238$ and 240
are not shown in the figure, as they are rather 
similar
to the ones obtained for nuclei with $A=236$ in each isotopic 
chain. As can be seen from the figure, all the PESs exhibit 
a similar topography, i.e., steepness along both the $\beta_2$
and $\beta_4$ directions as well as the locations of the absolute
minima. 

In panels (a), (c) and (e) of Fig.~\ref{fig:beta_min} we have plotted the 
quadrupole $\beta_2^{\textnormal{min}}$ and hexadecapole 
$\beta_4^{\textnormal{min}}$ deformations corresponding 
to the absolute minima of the HFB PESs in Fig.~\ref{GOGNY_PES}. In panels
(b), (d) and (f) of the same figure, we have also plotted 
the $\beta_2^{\textnormal{min}}$
values along the hexadecapole $\beta_4=0$ path.

As can be seen from panels (a), (c) and (e), all 
the isotopes display pronounced quadrupole $\beta^{\textnormal{min}}_2>0$
deformations \cite{guzman2025}. For U and Pu
isotopes, the 
$\beta^{\textnormal{min}}_2$ deformations only change slightly
with the mass number $A$, starting with $\beta_2^{\textnormal{min}}=0.26$
at $A=232$, and increasing up to
$\beta_2^{\textnormal{min}}=0.28$
around $A=240$. In the case of Th isotopes, the largest quadrupole 
deformation $\beta_2^{\textnormal{min}}=0.28$
corresponds to $^{238}$Th, while for $^{240}$Th we have obtained 
the value $\beta_2^{\textnormal{min}}=0.26$.

For the considered Th, U and Pu nuclei, the Gogny-D1S HFB 
ground states correspond to 
diamond-like 
$\beta^{\textnormal{min}}_4>0$ 
shapes \cite{guzman2025}. The 
$\beta^{\textnormal{min}}_4$ 
values decrease 
with increasing $A$. Among the studied 
isotopic chains, the most pronounced 
decrease is observed for Th nuclei. On the 
other hand, for Pu isotopes there is 
a slight increase from
$\beta_4^{\textnormal{min}}=0.18$ in $^{234}$Pu
to $\beta_4^{\textnormal{min}}=0.2$ in $^{236}$Pu,
after which the $\beta_4^{\textnormal{min}}$ values
continue decreasing.

As can be seen from panels (b), (d) and (f), the 
$\beta_2^{\textnormal{min}}$ values along the 
$\beta_4=0$ path in the PESs are smaller than those 
found in the $(\beta_2,\beta_4)$-constrained 
HFB calculations \cite{guzman2025}. In this case, we have 
found $\beta_2^{\textnormal{min}}$ deformations  ranging from 
$\beta_2^{\textnormal{min}}=0.16$ ($^{232}$Pu)
or $0.18$ ($^{232}$Th and $^{232}$U)
to $\beta_2^{\textnormal{min}}=0.22$ ($^{240}$U)
or $0.24$ ($^{240}$Th and $^{240}$Pu).

The mapped $sdg$-IBM PESs obtained for
$^{232,234,236}$Th, $^{232,234,236}$U and 
$^{232,234,236}$Pu are shown in Fig.~\ref{IBM_PES}. From
the comparison with the fermionic PESs in 
Fig.~\ref{GOGNY_PES}, one realizes that the mapping 
procedure reproduces the location of the absolute 
minima. Nevertheless, the  $sdg$-IBM PESs are flatter
than the SCMF ones in the region corresponding to
larger $\beta_2$ and $\beta_{4}$
deformations. This difference between the mapped IBM and SCMF
PESs is a well known feature arising  from the different size of 
the models spaces employed in each of these frameworks. The IBM
model considers a given valence space, while 
all the nucleonic  degrees of freedom are incorporated 
in a much larger fermionic
model space within the SCMF scheme.

\begin{figure*}    
\begin{center}
\includegraphics[width=\linewidth]{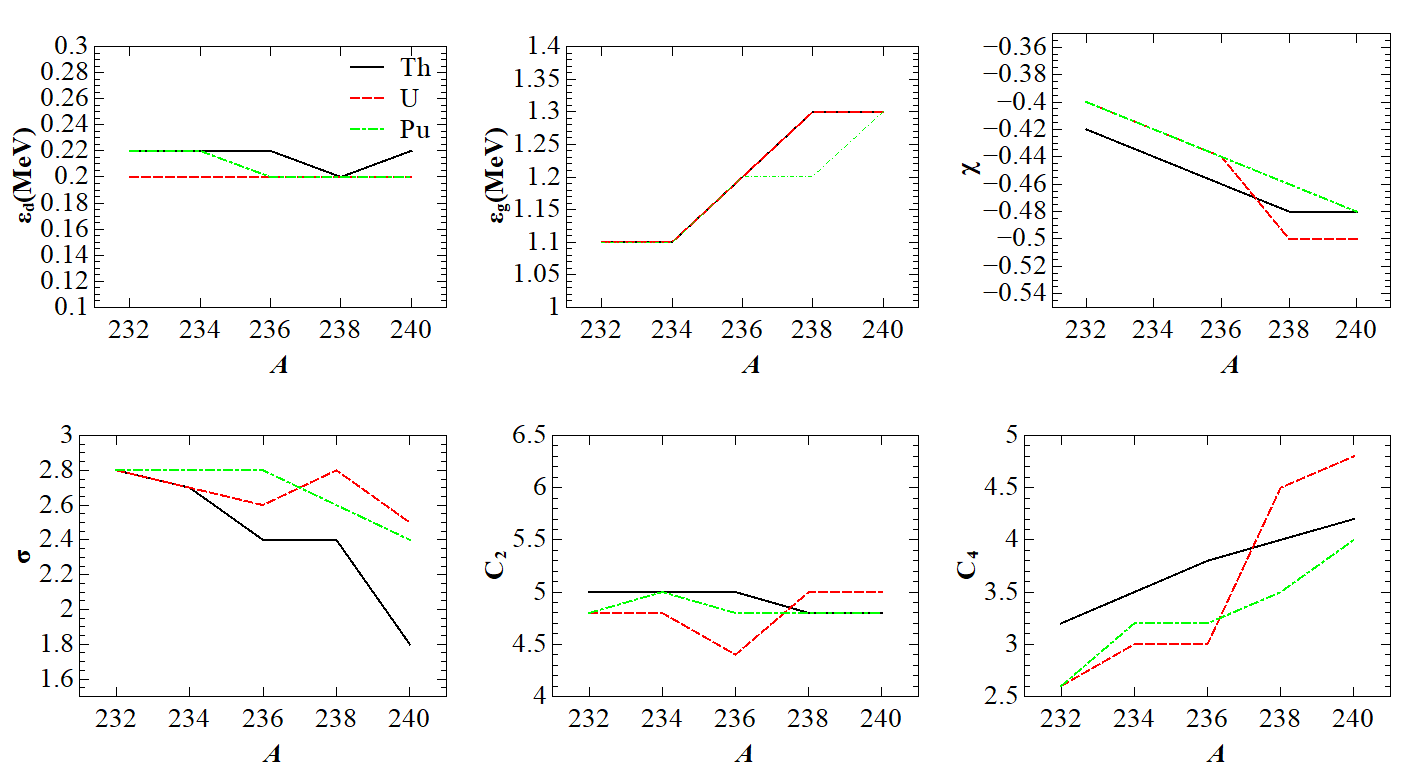}
\caption{(Color online) The parameters of the
$sdg$-IBM Hamiltonian Eq.~(\ref{eq:hb_sdg}), obtained for 
 $^{232-240}$Th,  $^{232-240}$U and  $^{232-240}$Pu, 
are plotted as functions of the mass number  $A$. For more details, see 
the main text.} 
\label{fig:sdgpar}
\end{center}
\end{figure*}

\begin{figure*}    
\begin{center}
\includegraphics[width =0.8\linewidth]{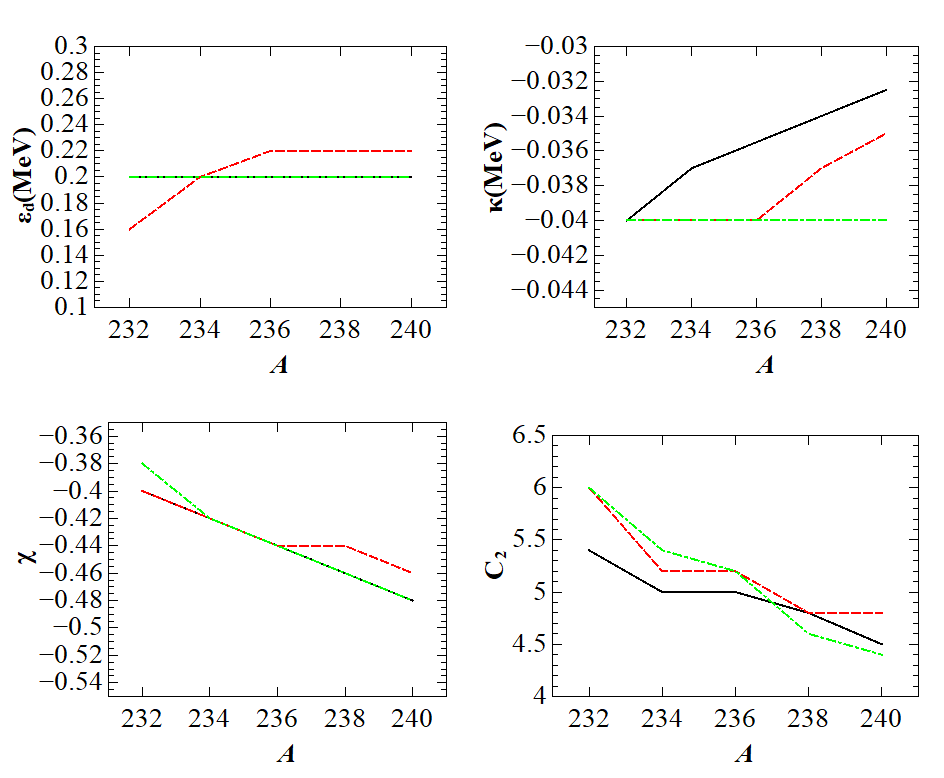}
\caption{(Color online) The parameters of the
$sd$-IBM Hamiltonian Eq.~\eqref{eq:hb_sd}, obtained for 
 $^{232-240}$Th,  $^{232-240}$U and  $^{232-240}$Pu, 
are plotted as functions of the mass number  $A$. For more details, see 
the main text.}
\label{fig:sdpar}
\end{center}
\end{figure*}

The $sdg$- and $sd$-IBM parameters, obtained via 
the mapping procedure, are
depicted in Figs. \ref{fig:sdgpar}
and \ref{fig:sdpar}, as functions of the mass number 
$A$. In the $sdg$-IBM, the $d$-boson energy remains
nearly constant at
$\epsilon_d\approx0.2-0.22$ MeV. This reflects, that the
quadrupole deformations $\beta_{2}^{\rm min}$
do not change significantly with $A$. In the $sd$-IBM, $\epsilon_d$
stays constant at 0.2 MeV for 
Th and Pu isotopes, while for 
U isotopes, it slightly increases
with $A$.

Within the $sdg$-IBM, the parameter $\kappa$, which is not shown in
Fig.~\ref{fig:sdgpar},  is constant ($\kappa=-0.04$ MeV). This is because, for 
the considered nuclei, the topography of the PESs is
similar and the location of the absolute minimum 
does not differ much from one another. However, within the
$sd$-IBM, the parameter $\kappa$ (see, Fig.~\ref{fig:sdpar}) is constant for Pu isotopes, while 
for Th and U nuclei it slightly decreases in magnitude
with $A$.

The $\chi$ parameter exhibits a similar
behavior in both the mapped  $sdg$-IBM
and $sd$-IBM  models, i.e., it increases 
its absolute value with $A$. This reflects
the increase of the $\beta_2^{\textnormal{min}}$ values.
However, the parameter $C_2$
exhibits a different behavior in the two IBM models. 
In the case of the $sdg$-IBM, it fluctuates between
 $C_2=5.0$ and 4.5, while it decreases
monotonously with $A$ in the $sd$-IBM. This different
behavior results from the fact that in the
$(\beta_2,\beta_4)$-constrained Gogny-D1S
HFB calculations the $\beta_2^{\textnormal{min}}$
values slightly increase with $A$. On the other 
hand, the quadrupole deformation corresponding to the
minimum along
the $\beta_4=0$ axis increases
more significantly with $A$
(see Fig.~\ref{fig:beta_min}).

The parameters $\epsilon_g$ (the $g$-boson energy), $\sigma$
and $C_4$ are only relevant within the mapped $sdg$-IBM (see, Fig.~\ref{fig:sdgpar}). The values
of $\epsilon_g$ increase, as functions of $A$. This reflects, the 
decreasing trend observed in the $\beta_4^{\textnormal{min}}$ values
in panels (a), (c) and (e) of Fig.~\ref{fig:beta_min}. The 
parameter $\sigma$ decreases with $A$, which corresponds to  the
decrease of both $\beta_4^{\textnormal{min}}$ and the ratio
$\beta_4^{\textnormal{min}}/\beta_2^{\textnormal{min}}$.
Finally, the increase of $C_4$
with $A$ reflects the decreasing pattern
of $\beta_4^{\textnormal{min}}$ with $A$.

\section{Results of the spectroscopic calculations\label{sec:results}}

\subsection{Excitation energies}

\begin{figure*}
\begin{center}
\includegraphics[width =.7\linewidth]{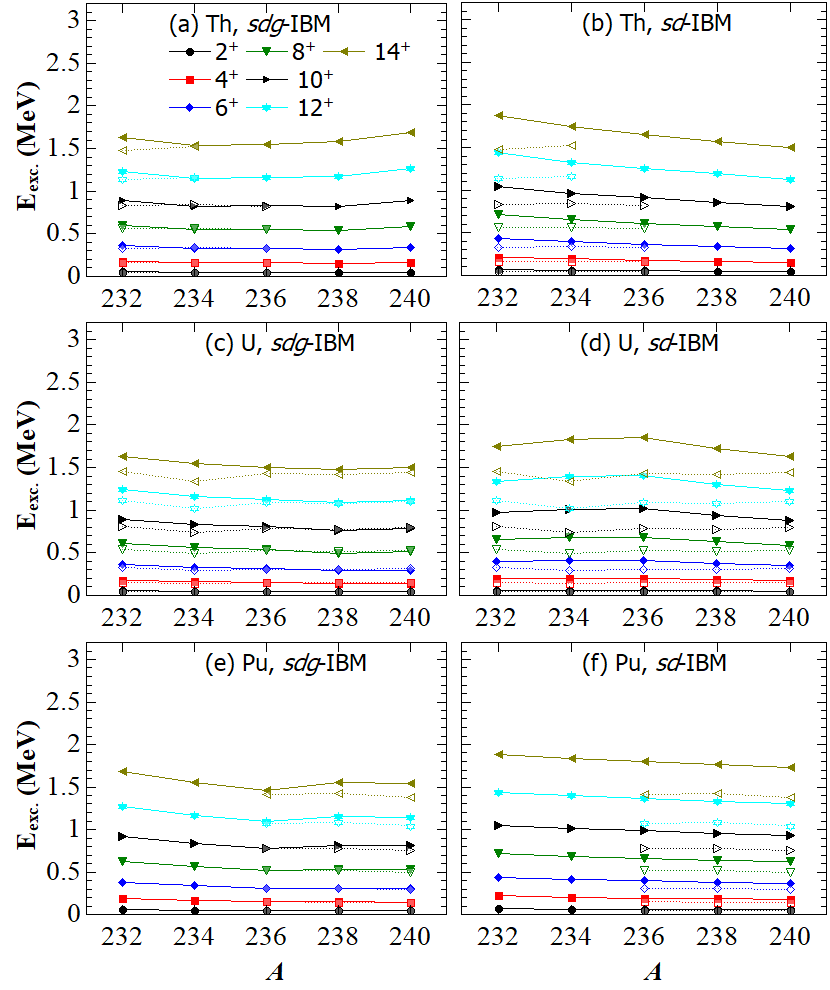}
\caption{(Color online) The excitation energies of the yrast band states with spins
from $J^{\pi}=2^+$ up to $J^{\pi}=14^+$ (solid symbols
connected by the solid lines), obtained within the $sdg$-IBM
model for $^{232-240}$Th, $^{232-240}$U and $^{232-240}$Pu, are plotted in panels (a), (c) and (e) as functions 
of the mass number $A$. The corresponding 
excitation energies, obtained within the $sd$-IBM model, ar shown 
in panels (b), (d) and (f). The available experimental data, taken 
from Ref.~\cite{data}, are depicted as open symbols connected by 
the dotted lines.}
\label{yrast}
\end{center}
\end{figure*}

The excitation energies of the yrast  states with spins in the range
 $J^{\pi}=2^+-14^+$, obtained within the $sdg$-IBM
model, are plotted in panels (a), (c) and (e)
of Fig.~\ref{yrast}, as functions 
of the mass number $A$. The 
excitation energies, obtained within the $sd$-IBM model, are shown 
in panels (b), (d) and (f) of the same figure. One 
realizes that the $sdg$-IBM improves the description of the excitation energies
of the 
high-spin  yrast states with $J^+ \geq 10^+$ in almost all of the studied isotopes, as compared 
with the results obtained within the $sd$-IBM. For the  $A=232$ and 234 isobars, both 
IBM models tend to overestimate the experimental excitation energies 
\cite{data}
of those states. However, the $sdg$-IBM energies are much closer to the 
experiment than the $sd$-IBM ones. This reduction of the 
excitation energies of higher-spin states represents 
a major
outcome of considering
the hexadecapole degree
of freedom in the $sdg$-IBM calculations.

The $g$-boson content in a given state
can be studied via
the expectation value of the
$g$-boson number
operator $\braket{\hat{n}_g}$ computed 
with the corresponding
$sdg$-IBM wave function. For the yrast states
of nuclei
with $A=232-236$, we have obtained 
$0.6 \lesssim \left < \hat{n}_g \right > \lesssim 0.8$. On the other 
hand, for those states in nuclei with $A=238,240$ we have obtained 
average values  $\braket{\hat{n}_g}\approx 0.4$. From these results, we 
conclude that, to a certain extent, hexadecapole correlations play a role 
in the yrast bands  of actinide nuclei. However, the predicted 
$\braket{\hat n_g}$ values are much smaller than the ones found 
in our previous mapped $sdg$-IBM calculations \cite{lotina2025} for rare-earth isotopes, based
on the parametrization D1S of the Gogny-EDF.

The improvements in the excitation energies of the
yrast states within the $sdg$-IBM, with respect to the $sd$-IBM, are 
also
significantly less pronounced
than those obtained for rare-earth nuclei. The 
inclusion of the $g$ boson in the $sdg$-IBM calculations leads to 
a lowering in the excitation energies of only  $\sim 0.3$ MeV for 
the considered 
actinide nuclei, while for  rare-earth nuclei the
$g$ boson lowers the energy levels
more significantly \cite{lotina2025}. This suggests
that $g$ bosons have a more limited impact 
in the higher-spin yrast states of actinides, as compared 
with rare-earth nuclei.
In order to understand the discrepancies it is worth mentioning that the
Gogny-D1S PESs for the studied Th, U and Pu nuclei
are not as soft along the $\beta_4$ direction
as those obtained in the rare-earth region
\cite{lotina2024hex-2,lotina2025,guzman2025,guzman2025rare_earth}

\begin{figure*}
\begin{center}
\includegraphics[width=.7\linewidth]{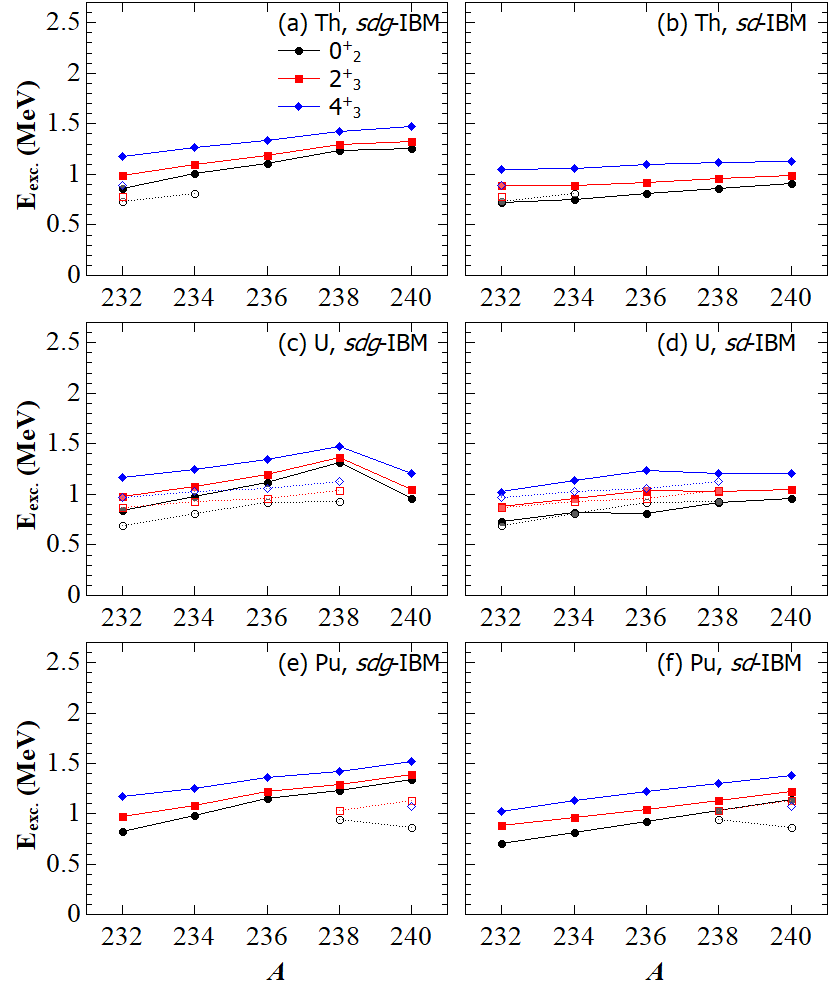}
\caption{(Color online) The excitation energies of the $0_{2}^+$, $2_{3}^+$, and $4_{3}^+$ states
(solid symbols
connected by the solid lines), obtained within the $sdg$-IBM
model for $^{232-240}$Th, $^{232-240}$U and $^{232-240}$Pu, are plotted in panels (a), (c) and (e) as functions 
of the mass number $A$. The corresponding 
excitation energies, obtained within the $sd$-IBM model, ar shown 
in panels (b), (d) and (f). The available experimental data, taken 
from Ref.~\cite{data}, are depicted as open symbols connected by 
the dotted lines.} 
\label{fig:0+}
\end{center}
\end{figure*}

In this study, the nonyrast $0^+_2$, $2^+_3$ and $4^+_3$
states have been considered as members of
a possible $K^{\pi}=0^+_2$ band according to 
their level ordering
and the dominant $\Delta J=2$
$E2$ transitions between them. The excitation energies 
predicted for these states are plotted in Fig.~\ref{fig:0+}, as
functions of the mass number $A$. From the comparison between the 
$sdg$-IBM results [panels (a), (c) and (e)]  with 
the $sd$-IBM ones [panels (b), (d) and (f)], we conclude that
the former model does not provide an improvement 
with respect to the latter. Note, that for the 
$A=238$ isobars the $sdg$-IBM even overestimates
the energies of
these nonyrast states. It is possible that the use of a more
general $sdg$-IBM Hamiltonian \cite{kuyucak1994} could 
improve the description of the nonyrast states. Nevertheless, one
should also keep in mind that the expectation values
$\left < \hat{n}_g \right >$ are 
significantly small, indicating that 
hexadecapole correlations are rather
minor in those states. The study of the hexadecapole effects
in actinides with a more general $sdg$-IBM
Hamiltonian is out of the scope of this paper. Work along these 
lines will be reported in future publications.


\begin{figure*}
\begin{center}
\includegraphics[width=.7\linewidth]{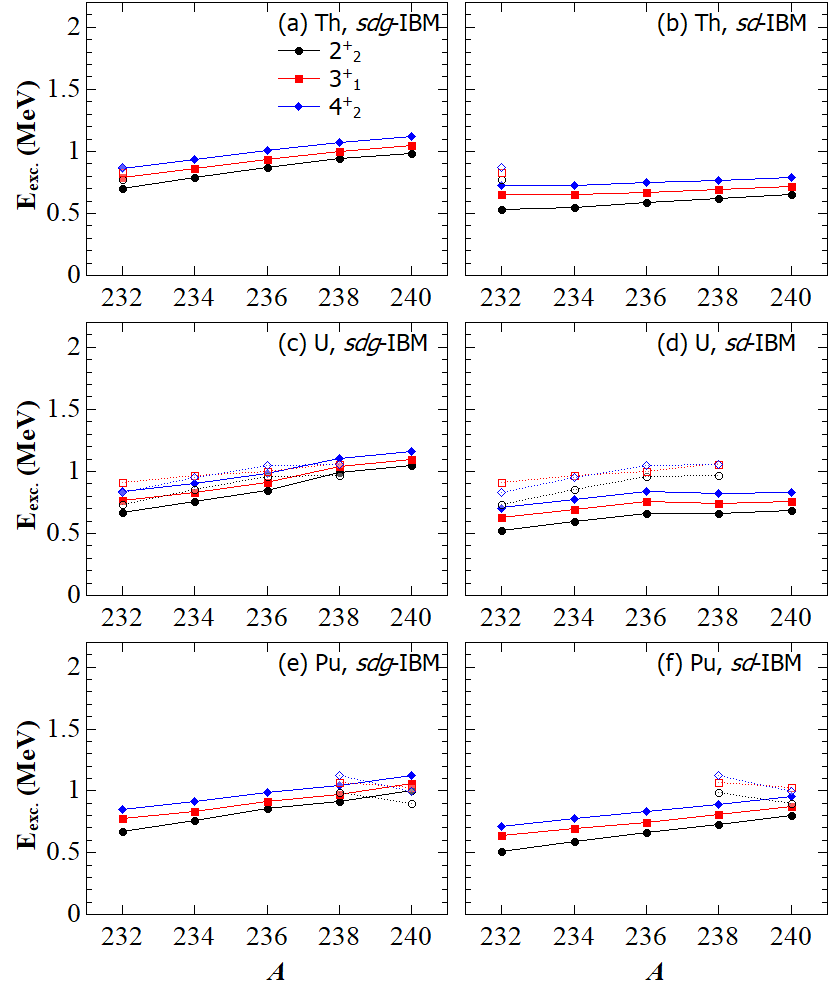}
\caption{(Color online) The excitation energies of the  $2_{2}^+$, $3_{1}^+$, and $4_{2}^+$ states
(solid symbols
connected by the solid lines), obtained within the $sdg$-IBM
model for $^{232-240}$Th, $^{232-240}$U and $^{232-240}$Pu, are plotted in panels (a), (c) and (e) as functions 
of the mass number $A$. The corresponding 
excitation energies, obtained within the $sd$-IBM model, ar shown 
in panels (b), (d) and (f). The available experimental data, taken 
from Ref.~\cite{data}, are depicted as open symbols connected by 
the dotted lines.} 
\label{gamma}
\end{center}
\end{figure*}

The excitation energies of the $2^+_2$, $3^+_1$, and $4^+_2$ states, 
obtained within the $sdg$-IBM model, are plotted in panels (a), (c) and 
(e) of Fig.~\ref{gamma}, as functions of the mass number $A$. The 
excitation energies, obtained within the $sd$-IBM model, are shown in 
panels (b), (d) and (f) of the same figure. Similar to the $K^\pi=0^+$ 
excited band, we have associated the $2^+_2$, $3^+_1$, and $4^+_2$ 
states with a possible $K^\pi=2^+$ or $\gamma$-vibrational band 
according to the level structure and $E2$ transition patterns. The 
$sdg$-IBM calculations provide larger excitation energies for the 
$2^+_2$, $3^+_1$, and $4^+_2$ states than in the $sd$-IBM case and 
gives a slight improvement in describing the experimental low-lying 
structure in most of the studied nuclei. The $g$-boson expectation 
values in these states are $\braket{\hat{n}_g} \approx 0.4$, which 
implies some admixture of hexadecapole components. However, none of the 
considered mapped IBM models includes triaxiality, i.e., the $\gamma$ 
degree of freedom that might improve the description of the 
$\gamma$-band states \cite{nomura2012tri} and therefore it is rather 
unclear whether the improvement in the comparison with experiment of 
the members of this band can be ascribed to the hexadecapole 
collectivity in actinides. Triaxiality can be included in both the 
$sdg$-IBM and $sd$-IBM \cite{vanisacker2010}, as well as in the SCMF 
calculations \cite{RAY-gamma-1,RAY-gamma-2,lu2014}. However, such an 
extension is out of the scope of this work as it would lead to 
multi-dimensional constrained SCMF calculations, which would 
significantly complicate the mapping procedure.

\subsection{Transition strengths}

\subsubsection{$E2$ transitions}

\begin{figure}
\begin{center}
\includegraphics[width=.8\linewidth]{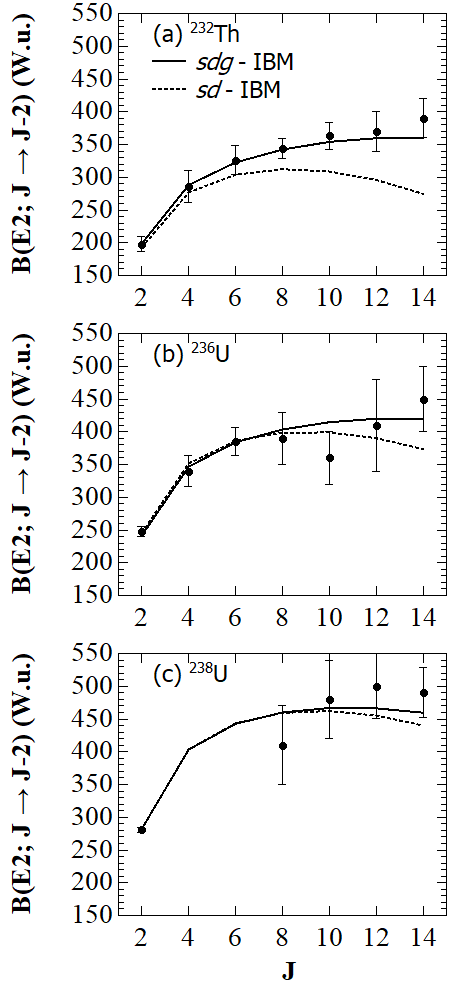}
\caption{$B(E2)$ values 
along the yrast bands
of $^{232}$Th [panel (a)], $^{236}$U [panel (b)] and $^{238}$U [panel (c)]. 
The solid lines represent the results
obtained within the $sdg$-IBM, while
the dotted lines connect
the $sd$-IBM results.
The experimental data, represented by the
solid circles, 
are taken from Ref.~\cite{data}.} 
\label{E2}
\end{center}
\end{figure}

The $B(E2; J \rightarrow J-2)$ 
transition strengths along the yrast bands of
$^{232}$Th, $^{236}$U and $^{238}$U are shown in 
panels (a), (b) and (c) of Fig.~\ref{E2}. These
nuclei have been chosen as illustrative examples for 
which, experimental data on $E2$ transition
strengths for yrast states
with $J^{\pi}= 2^+-14^+$ are available. For $^{232}$Th, the
$sdg$-IBM significantly
improves the description of the $E2$ transition
strengths over the $sd$-IBM predictions for
yrast states with $J^{\pi}\geqslant 6^+$.
However, for both $^{236}$U and $^{238}$U,
an improvement is only obtained
in the $J^{\pi}=12^+$ and $14^+$ states.
This reflects that
the calculated
$\braket{\hat{n}_g}$ values
are larger in the yrast states of 
nuclei with $A=232$ and 234, as
compared with the  $\braket{\hat{n}_g}$ values
obtained for $A\geqslant 236$. As a result, the 
$\sigma (d^{\dagger} \times \tilde{g} + g^{\dagger} \times \tilde{g})^{(2)}$
term in the quadrupole operator Eq.~(\ref{eq3}), which is also
part of the $\hat{T}^{(E2)}$ transition
operator in Eq.~(\ref{eq9}), contributes 
less to the calculated $B(E2)$ strengths along the 
yrast bands of nuclei with $A\geqslant 236$.

\subsubsection{Hexadecapole transitions}

\begin{figure}
\begin{center}
\includegraphics[width=.8\linewidth]{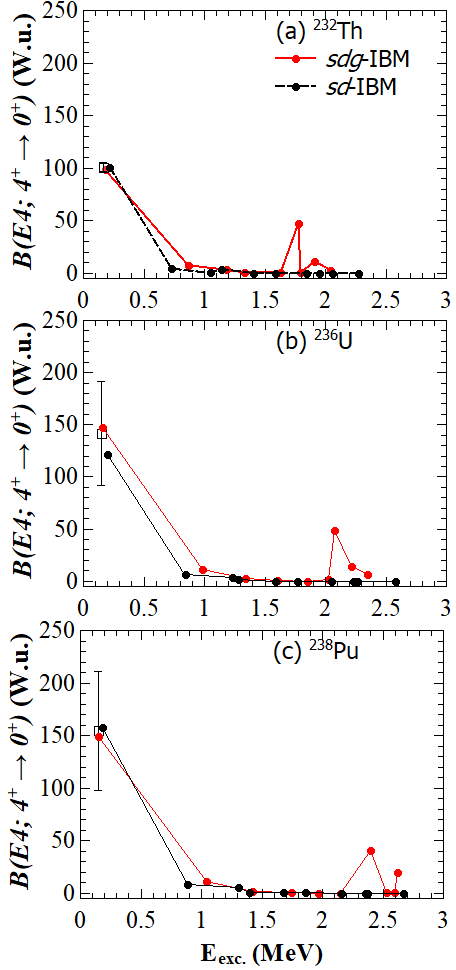}
\caption{(Color online) The $sdg$-IBM (red) 
and $sd$-IBM (black) $B(E4; 4^+_n \to 0^+_1)$ ($n=1,2,...,10$)
values, obtained for the nuclei $^{232}$Th [panel (a)], $^{236}$U [panel (b)] 
and $^{238}$Pu [panel (c)], are
depicted as functions of the excitation energies
of the $4^+_n$ states. The experimental $B(E4; 4^+_1 \rightarrow 0^+_1)$ data,
indicated by open symbols, are taken from Ref.~\cite{data}.}
\label{fig:E4}
\end{center}
\end{figure}

The  $B(E4; 4^+_n \to 0^+_1)$ ($n=1,2,...,10$) values, obtained for the 
nuclei $^{232}$Th, $^{236}$U and $^{238}$Pu, and within the $sdg$-IBM 
and $sd$-IBM model spaces are shown in panels (a), (b) and (c) of 
Fig.~\ref{fig:E4}, as functions of the excitation energies of the 
$4^+_n$ states. Those reduced transition probabilities have been 
computed using the $E4$ operators \eqref{eq10} for the $sdg$-IBM and 
\eqref{eq11} for the $sd$-IBM. We have considered the nuclei 
$^{232}$Th, $^{236}$U and $^{238}$Pu as illustrative examples for 
which, experimental data for the $B(E4; 4^+_1 \to 0^+_1)$ strengths  
are available. The $sdg$-IBM and $sd$-IBM $B(E4; 4^+_1 \rightarrow 
0^+_1)$ values are similar, as the the $e_4$ effective charges are 
fitted to reproduce the measured hexadecapole transition strengths. One 
should note, however, that the $e_4$ charges in the $sd$-IBM are 
significantly larger than the $sdg$-IBM ones. This points towards the 
need of considering $g$-bosons when computing  $B(E4)$ transition 
strengths. Besides the first $4^+$ state, the $sd$-IBM does not predict 
any strong $E4$ transitions from higher-lying $4^+$ states to the 
ground state. However, the mapped $sdg$-IBM model predicts relatively 
large $B(E4)$ values for the $4^+_7$ state ($^{232}$Th and $^{238}$Pu) 
or the $4^+_8$ state ($^{236}$U). These $4^+_7$ and/or $4^+_8$ states 
can be considered to be bandheads of the $K^\pi=4^+$ bands since for 
these states, the $sdg$-IBM provides  expectation values $\left < 
\hat{n}_g \right > \approx 1$.

For some rare-earth nuclei, considered in previous studies 
\cite{lotina2024hex-2,lotina2025}, the $K^\pi=4^+$ bandhead  $4^+_3$ or 
$4^+_4$ states (containing one $g$ boson) were predicted at much lower 
excitation energies ($E_{\rm exc} < 2.0$ MeV). The fact that the first 
$4^+$ states, in which $g$ bosons are present, are significantly higher 
in energy in actinides, indicates that hexadecapole correlations are 
weaker in actinides than in  rare-earth nuclei. Therefore, hexadecapole 
correlations are more significantly overshadowed by quadrupole effects 
in  actinides. There is no experimental information available on the 
$4^+_7$ and $4^+_8$ states. Thus, at the present stage, it is not 
possible to assess the accuracy of the mapped $sdg$-IBM regarding the 
properties of the $4^+$ bands in actinides.


\section{Summary\label{sec:summary}}

We have carried out a detailed analysis of the quadrupole-hexadecapole 
collectivity in the even-even actinide nuclei $^{232-240}$Th, 
$^{232-240}$U, and $^{232-240}$Pu where hexadecapole correlations are 
expected to play a role in the low-energy dynamics. To this end, we 
have resorted to the Gogny-D1S HFB framework as initial (fermionic) 
input followed by the mapped $sdg$-IBM. We have also performed 
comparisons with results obtained within the simpler mapped $sd$-IBM 
model in order to understand the role played by the $g$ (hexadecapole) 
boson.

For most of the studied nuclei, the minima of the 
$(\beta_2,\beta_4)$-PESs correspond to quadrupole deformations 
$\beta_2\approx0.25$ and hexadecapole deformations $\beta_{4}>0$. The  
excitation energies of low-lying positive-parity yrast and nonyrast 
states, obtained via the diagonalization of the $sdg$-IBM Hamiltonian, 
agree reasonably well with the available experimental data. The results 
obtained in this work and in the previous study \cite{lotina2025} show, 
that the Gogny-HFB mapped $sdg$-IBM model represents a reasonable 
starting point to describe the spectroscopic properties of nuclei in 
different regions of the nuclear chart, such as rare-earth and actinide 
nuclei for which hexadecapole collectivity  plays a role.

We have shown that the mapped $sdg$-IBM model can improve the 
description of the high-spin yrast states in almost all the studied 
nuclei, especially those with mass numbers $A=232$ and 234, for which 
hexadecapole collectivity is predicted to play a role  in the yrast 
states with even spins $J^{\pi}\geqslant 10^+$. As compared with the 
rare-earth region, hexadecapole effects have been found to be rather 
weak in actinides. For nonyrast states, members of the excited 
$K^{\pi}=0^+$ and $K^{\pi}=2^+$ bands, the $sdg$-IBM overestimates the 
energies of the excited $0^+$ states and provides minor improvements in 
the excitation energies of states in the $K^\pi=2^+$ band, as compared 
with the results obtained with the simpler $sd$-IBM Hamiltonian. A more 
general $sdg$-IBM Hamiltonian could be necessary to better understand 
how the inclusion of the $g$-boson affects the energies of the states 
in the non-yrast bands.

In comparison with the $sd$-IBM, the $sdg$-IBM predicts larger $B(E2)$ 
values between yrast states, especially for $A=232$ isobars. This 
agrees well with the available experimental data and suggests that 
hexadecapole correlations represent a necessary  building block to 
account for  $E2$ transitions in the yrast band of nuclei in this 
region.

For larger mass numbers, in each isotopic chain, the differences 
between the $sd$-IBM and $sdg$-IBM $B(E2)$ strengths become smaller, 
though the effects of hexadecapole collectivity can still be observed 
in the $J^{\pi}=14^+$ states. The $sdg$-IBM model predicts 
$K^{\pi}=4^+$ bands with enhanced $4^+ \rightarrow 0^+$ $E4$ 
transitions. However, the $4^+_7$ and/or $4^+_8$ states are predicted 
at higher excitation energies than the $K^{\pi}=4^+$ bandheads found 
for rare-earth nuclei \cite{lotina2024hex-2,lotina2025}. Experimental 
information on $E4$ transitions in actinides is required to better 
understand to which extent the $sdg$-IBM model accounts for the 
properties of the $K^{\pi}=4^+$ bands.

A long list of tasks remains to be done. For example, a potential 
extension of this study would be to apply the Gogny-HFB mapped 
$sdg$-IBM framework  to regions of the nuclear chart characterized by 
negative hexadecapole deformations, i.e., square-like shapes, such as W 
and Os \cite{guzman2025rare_earth} as well as Pt nuclei. Another 
possible extension of the present work could be the inclusion of both 
neutron and proton boson degrees of freedom to study the hexadecapole 
collectivity and mixed-symmetry states of nuclei.

\acknowledgments
The work of K. Nomura has been supported by JSPS KAKENHI Grant No. 
JP25K07293. The work of R. Rodr\'{\i}guez-Guzm\'an  is funded by 
Nazarbayev University under the Faculty Development Competitive 
Research Grants Program (FDCRGP) for 2025-2027, Grant 040225FD4712. The 
work of L. M. Robledo is supported by Spanish Agencia Estatal de 
Investigacion (AEI) of the Ministry of Science and Innovation under 
Grant No. PID2021-127890NB-I00 and PID2024-159559NB-C21.

\bibliography{refs}

\begin{thebibliography}{66}%
\makeatletter
\providecommand \@ifxundefined [1]{%
 \@ifx{#1\undefined}
}%
\providecommand \@ifnum [1]{%
 \ifnum #1\expandafter \@firstoftwo
 \else \expandafter \@secondoftwo
 \fi
}%
\providecommand \@ifx [1]{%
 \ifx #1\expandafter \@firstoftwo
 \else \expandafter \@secondoftwo
 \fi
}%
\providecommand \natexlab [1]{#1}%
\providecommand \enquote  [1]{``#1''}%
\providecommand \bibnamefont  [1]{#1}%
\providecommand \bibfnamefont [1]{#1}%
\providecommand \citenamefont [1]{#1}%
\providecommand \href@noop [0]{\@secondoftwo}%
\providecommand \href [0]{\begingroup \@sanitize@url \@href}%
\providecommand \@href[1]{\@@startlink{#1}\@@href}%
\providecommand \@@href[1]{\endgroup#1\@@endlink}%
\providecommand \@sanitize@url [0]{\catcode `\\12\catcode `\$12\catcode
  `\&12\catcode `\#12\catcode `\^12\catcode `\_12\catcode `\%12\relax}%
\providecommand \@@startlink[1]{}%
\providecommand \@@endlink[0]{}%
\providecommand \url  [0]{\begingroup\@sanitize@url \@url }%
\providecommand \@url [1]{\endgroup\@href {#1}{\urlprefix }}%
\providecommand \urlprefix  [0]{URL }%
\providecommand \Eprint [0]{\href }%
\providecommand \doibase [0]{https://doi.org/}%
\providecommand \selectlanguage [0]{\@gobble}%
\providecommand \bibinfo  [0]{\@secondoftwo}%
\providecommand \bibfield  [0]{\@secondoftwo}%
\providecommand \translation [1]{[#1]}%
\providecommand \BibitemOpen [0]{}%
\providecommand \bibitemStop [0]{}%
\providecommand \bibitemNoStop [0]{.\EOS\space}%
\providecommand \EOS [0]{\spacefactor3000\relax}%
\providecommand \BibitemShut  [1]{\csname bibitem#1\endcsname}%
\let\auto@bib@innerbib\@empty
\bibitem [{\citenamefont {Bohr}\ and\ \citenamefont {Mottelson}(1975)}]{BM}%
  \BibitemOpen
  \bibfield  {author} {\bibinfo {author} {\bibfnamefont {A.}~\bibnamefont
  {Bohr}}\ and\ \bibinfo {author} {\bibfnamefont {B.~R.}\ \bibnamefont
  {Mottelson}},\ }\href@noop {} {\emph {\bibinfo {title} {Nuclear
  Structure}}},\ Vol.~\bibinfo {volume} {II}\ (\bibinfo  {publisher} {Benjamin,
  New York, USA},\ \bibinfo {year} {1975})\BibitemShut {NoStop}%
\bibitem [{\citenamefont {Gupta}\ \emph {et~al.}(2020)\citenamefont {Gupta},
  \citenamefont {Nayak}, \citenamefont {Garg}, \citenamefont {Hagino},
  \citenamefont {Howard}, \citenamefont {Sensharma}, \citenamefont
  {Şenyiğit}, \citenamefont {Tan}, \citenamefont {O'Malley}, \citenamefont
  {Smith}, \citenamefont {Gandhi}, \citenamefont {Anderson}, \citenamefont
  {deBoer}, \citenamefont {Frentz}, \citenamefont {Gyurjinyan}, \citenamefont
  {Hall}, \citenamefont {Hall}, \citenamefont {Hu}, \citenamefont {Lamere},
  \citenamefont {Liu}, \citenamefont {Long}, \citenamefont {Lu}, \citenamefont
  {Lyons}, \citenamefont {Ostdiek}, \citenamefont {Seymour}, \citenamefont
  {Skulski},\ and\ \citenamefont {{Vande Kolk}}}]{gupta2020}%
  \BibitemOpen
  \bibfield  {author} {\bibinfo {author} {\bibfnamefont {Y.}~\bibnamefont
  {Gupta}}, \bibinfo {author} {\bibfnamefont {B.}~\bibnamefont {Nayak}},
  \bibinfo {author} {\bibfnamefont {U.}~\bibnamefont {Garg}}, \bibinfo {author}
  {\bibfnamefont {K.}~\bibnamefont {Hagino}}, \bibinfo {author} {\bibfnamefont
  {K.}~\bibnamefont {Howard}}, \bibinfo {author} {\bibfnamefont
  {N.}~\bibnamefont {Sensharma}}, \bibinfo {author} {\bibfnamefont
  {M.}~\bibnamefont {Şenyiğit}}, \bibinfo {author} {\bibfnamefont
  {W.}~\bibnamefont {Tan}}, \bibinfo {author} {\bibfnamefont {P.}~\bibnamefont
  {O'Malley}}, \bibinfo {author} {\bibfnamefont {M.}~\bibnamefont {Smith}},
  \bibinfo {author} {\bibfnamefont {R.}~\bibnamefont {Gandhi}}, \bibinfo
  {author} {\bibfnamefont {T.}~\bibnamefont {Anderson}}, \bibinfo {author}
  {\bibfnamefont {R.}~\bibnamefont {deBoer}}, \bibinfo {author} {\bibfnamefont
  {B.}~\bibnamefont {Frentz}}, \bibinfo {author} {\bibfnamefont
  {A.}~\bibnamefont {Gyurjinyan}}, \bibinfo {author} {\bibfnamefont
  {O.}~\bibnamefont {Hall}}, \bibinfo {author} {\bibfnamefont {M.}~\bibnamefont
  {Hall}}, \bibinfo {author} {\bibfnamefont {J.}~\bibnamefont {Hu}}, \bibinfo
  {author} {\bibfnamefont {E.}~\bibnamefont {Lamere}}, \bibinfo {author}
  {\bibfnamefont {Q.}~\bibnamefont {Liu}}, \bibinfo {author} {\bibfnamefont
  {A.}~\bibnamefont {Long}}, \bibinfo {author} {\bibfnamefont {W.}~\bibnamefont
  {Lu}}, \bibinfo {author} {\bibfnamefont {S.}~\bibnamefont {Lyons}}, \bibinfo
  {author} {\bibfnamefont {K.}~\bibnamefont {Ostdiek}}, \bibinfo {author}
  {\bibfnamefont {C.}~\bibnamefont {Seymour}}, \bibinfo {author} {\bibfnamefont
  {M.}~\bibnamefont {Skulski}},\ and\ \bibinfo {author} {\bibfnamefont
  {B.}~\bibnamefont {{Vande Kolk}}},\ }\href
  {https://doi.org/https://doi.org/10.1016/j.physletb.2020.135473} {\bibfield
  {journal} {\bibinfo  {journal} {Phys. Lett. B}\ }\textbf {\bibinfo {volume}
  {806}},\ \bibinfo {pages} {135473} (\bibinfo {year} {2020})}\BibitemShut
  {NoStop}%
\bibitem [{\citenamefont {Spieker}\ \emph {et~al.}(2023)\citenamefont
  {Spieker}, \citenamefont {Agbemava}, \citenamefont {Bazin}, \citenamefont
  {Biswas}, \citenamefont {Cottle}, \citenamefont {Farris}, \citenamefont
  {Gade}, \citenamefont {Ginter}, \citenamefont {Giraud}, \citenamefont
  {Kemper}, \citenamefont {Li}, \citenamefont {Nazarewicz}, \citenamefont
  {Noji}, \citenamefont {Pereira}, \citenamefont {Riley}, \citenamefont
  {Smith}, \citenamefont {Weisshaar},\ and\ \citenamefont
  {Zegers}}]{spieker2023}%
  \BibitemOpen
  \bibfield  {author} {\bibinfo {author} {\bibfnamefont {M.}~\bibnamefont
  {Spieker}}, \bibinfo {author} {\bibfnamefont {S.}~\bibnamefont {Agbemava}},
  \bibinfo {author} {\bibfnamefont {D.}~\bibnamefont {Bazin}}, \bibinfo
  {author} {\bibfnamefont {S.}~\bibnamefont {Biswas}}, \bibinfo {author}
  {\bibfnamefont {P.}~\bibnamefont {Cottle}}, \bibinfo {author} {\bibfnamefont
  {P.}~\bibnamefont {Farris}}, \bibinfo {author} {\bibfnamefont
  {A.}~\bibnamefont {Gade}}, \bibinfo {author} {\bibfnamefont {T.}~\bibnamefont
  {Ginter}}, \bibinfo {author} {\bibfnamefont {S.}~\bibnamefont {Giraud}},
  \bibinfo {author} {\bibfnamefont {K.}~\bibnamefont {Kemper}}, \bibinfo
  {author} {\bibfnamefont {J.}~\bibnamefont {Li}}, \bibinfo {author}
  {\bibfnamefont {W.}~\bibnamefont {Nazarewicz}}, \bibinfo {author}
  {\bibfnamefont {S.}~\bibnamefont {Noji}}, \bibinfo {author} {\bibfnamefont
  {J.}~\bibnamefont {Pereira}}, \bibinfo {author} {\bibfnamefont
  {L.}~\bibnamefont {Riley}}, \bibinfo {author} {\bibfnamefont
  {M.}~\bibnamefont {Smith}}, \bibinfo {author} {\bibfnamefont
  {D.}~\bibnamefont {Weisshaar}},\ and\ \bibinfo {author} {\bibfnamefont
  {R.}~\bibnamefont {Zegers}},\ }\href
  {https://doi.org/https://doi.org/10.1016/j.physletb.2023.137932} {\bibfield
  {journal} {\bibinfo  {journal} {Physics Letters B}\ }\textbf {\bibinfo
  {volume} {841}},\ \bibinfo {pages} {137932} (\bibinfo {year}
  {2023})}\BibitemShut {NoStop}%
\bibitem [{\citenamefont {Erb}\ \emph {et~al.}(1972)\citenamefont {Erb},
  \citenamefont {Holden}, \citenamefont {Lee}, \citenamefont {Saladin},\ and\
  \citenamefont {Saylor}}]{erb1972}%
  \BibitemOpen
  \bibfield  {author} {\bibinfo {author} {\bibfnamefont {K.~A.}\ \bibnamefont
  {Erb}}, \bibinfo {author} {\bibfnamefont {J.~E.}\ \bibnamefont {Holden}},
  \bibinfo {author} {\bibfnamefont {I.~Y.}\ \bibnamefont {Lee}}, \bibinfo
  {author} {\bibfnamefont {J.~X.}\ \bibnamefont {Saladin}},\ and\ \bibinfo
  {author} {\bibfnamefont {T.~K.}\ \bibnamefont {Saylor}},\ }\href
  {https://doi.org/10.1103/PhysRevLett.29.1010} {\bibfield  {journal} {\bibinfo
   {journal} {Phys. Rev. Lett.}\ }\textbf {\bibinfo {volume} {29}},\ \bibinfo
  {pages} {1010} (\bibinfo {year} {1972})}\BibitemShut {NoStop}%
\bibitem [{\citenamefont {Wollersheim}\ and\ \citenamefont {{W.
  Elze}}(1977)}]{wollersheim1977}%
  \BibitemOpen
  \bibfield  {author} {\bibinfo {author} {\bibfnamefont {H.}~\bibnamefont
  {Wollersheim}}\ and\ \bibinfo {author} {\bibfnamefont {T.}~\bibnamefont {{W.
  Elze}}},\ }\href
  {https://doi.org/https://doi.org/10.1016/0375-9474(77)90186-5} {\bibfield
  {journal} {\bibinfo  {journal} {Nucl. Phys. A}\ }\textbf {\bibinfo {volume}
  {278}},\ \bibinfo {pages} {87} (\bibinfo {year} {1977})}\BibitemShut
  {NoStop}%
\bibitem [{\citenamefont {Baker}\ \emph {et~al.}(1989)\citenamefont {Baker},
  \citenamefont {Sethi}, \citenamefont {Penumetcha}, \citenamefont {Emery},
  \citenamefont {Jones}, \citenamefont {Grimm},\ and\ \citenamefont
  {Whiten}}]{baker1989}%
  \BibitemOpen
  \bibfield  {author} {\bibinfo {author} {\bibfnamefont {F.}~\bibnamefont
  {Baker}}, \bibinfo {author} {\bibfnamefont {A.}~\bibnamefont {Sethi}},
  \bibinfo {author} {\bibfnamefont {V.}~\bibnamefont {Penumetcha}}, \bibinfo
  {author} {\bibfnamefont {G.}~\bibnamefont {Emery}}, \bibinfo {author}
  {\bibfnamefont {W.}~\bibnamefont {Jones}}, \bibinfo {author} {\bibfnamefont
  {M.}~\bibnamefont {Grimm}},\ and\ \bibinfo {author} {\bibfnamefont
  {M.}~\bibnamefont {Whiten}},\ }\href
  {https://doi.org/https://doi.org/10.1016/0375-9474(89)90147-4} {\bibfield
  {journal} {\bibinfo  {journal} {Nuclear Physics A}\ }\textbf {\bibinfo
  {volume} {501}},\ \bibinfo {pages} {546} (\bibinfo {year}
  {1989})}\BibitemShut {NoStop}%
\bibitem [{\citenamefont {Sethi}\ \emph {et~al.}(1991)\citenamefont {Sethi},
  \citenamefont {Hintz}, \citenamefont {Mihailidis}, \citenamefont {Mack},
  \citenamefont {Gazzaly}, \citenamefont {Jones}, \citenamefont {Pauletta},
  \citenamefont {Santi},\ and\ \citenamefont {Goutte}}]{sethi1991}%
  \BibitemOpen
  \bibfield  {author} {\bibinfo {author} {\bibfnamefont {A.}~\bibnamefont
  {Sethi}}, \bibinfo {author} {\bibfnamefont {N.~M.}\ \bibnamefont {Hintz}},
  \bibinfo {author} {\bibfnamefont {D.~N.}\ \bibnamefont {Mihailidis}},
  \bibinfo {author} {\bibfnamefont {A.~M.}\ \bibnamefont {Mack}}, \bibinfo
  {author} {\bibfnamefont {M.}~\bibnamefont {Gazzaly}}, \bibinfo {author}
  {\bibfnamefont {K.~W.}\ \bibnamefont {Jones}}, \bibinfo {author}
  {\bibfnamefont {G.}~\bibnamefont {Pauletta}}, \bibinfo {author}
  {\bibfnamefont {L.}~\bibnamefont {Santi}},\ and\ \bibinfo {author}
  {\bibfnamefont {D.}~\bibnamefont {Goutte}},\ }\href
  {https://doi.org/10.1103/PhysRevC.44.700} {\bibfield  {journal} {\bibinfo
  {journal} {Phys. Rev. C}\ }\textbf {\bibinfo {volume} {44}},\ \bibinfo
  {pages} {700} (\bibinfo {year} {1991})}\BibitemShut {NoStop}%
\bibitem [{\citenamefont {Bemis}\ \emph {et~al.}(1973)\citenamefont {Bemis},
  \citenamefont {McGowan}, \citenamefont {Ford}, \citenamefont {Milner},
  \citenamefont {Stelson},\ and\ \citenamefont {Robinson}}]{bemis1973}%
  \BibitemOpen
  \bibfield  {author} {\bibinfo {author} {\bibfnamefont {C.~E.}\ \bibnamefont
  {Bemis}}, \bibinfo {author} {\bibfnamefont {F.~K.}\ \bibnamefont {McGowan}},
  \bibinfo {author} {\bibfnamefont {J.~L.~C.}\ \bibnamefont {Ford}}, \bibinfo
  {author} {\bibfnamefont {W.~T.}\ \bibnamefont {Milner}}, \bibinfo {author}
  {\bibfnamefont {P.~H.}\ \bibnamefont {Stelson}},\ and\ \bibinfo {author}
  {\bibfnamefont {R.~L.}\ \bibnamefont {Robinson}},\ }\href
  {https://doi.org/10.1103/PhysRevC.8.1466} {\bibfield  {journal} {\bibinfo
  {journal} {Phys. Rev. C}\ }\textbf {\bibinfo {volume} {8}},\ \bibinfo {pages}
  {1466} (\bibinfo {year} {1973})}\BibitemShut {NoStop}%
\bibitem [{\citenamefont {Zumbro}\ \emph {et~al.}(1984)\citenamefont {Zumbro},
  \citenamefont {Shera}, \citenamefont {Tanaka}, \citenamefont {Bemis},
  \citenamefont {Naumann}, \citenamefont {Hoehn}, \citenamefont {Reuter},\ and\
  \citenamefont {Steffen}}]{zumbro1984}%
  \BibitemOpen
  \bibfield  {author} {\bibinfo {author} {\bibfnamefont {J.~D.}\ \bibnamefont
  {Zumbro}}, \bibinfo {author} {\bibfnamefont {E.~B.}\ \bibnamefont {Shera}},
  \bibinfo {author} {\bibfnamefont {Y.}~\bibnamefont {Tanaka}}, \bibinfo
  {author} {\bibfnamefont {C.~E.}\ \bibnamefont {Bemis}}, \bibinfo {author}
  {\bibfnamefont {R.~A.}\ \bibnamefont {Naumann}}, \bibinfo {author}
  {\bibfnamefont {M.~V.}\ \bibnamefont {Hoehn}}, \bibinfo {author}
  {\bibfnamefont {W.}~\bibnamefont {Reuter}},\ and\ \bibinfo {author}
  {\bibfnamefont {R.~M.}\ \bibnamefont {Steffen}},\ }\href
  {https://doi.org/10.1103/PhysRevLett.53.1888} {\bibfield  {journal} {\bibinfo
   {journal} {Phys. Rev. Lett.}\ }\textbf {\bibinfo {volume} {53}},\ \bibinfo
  {pages} {1888} (\bibinfo {year} {1984})}\BibitemShut {NoStop}%
\bibitem [{\citenamefont {Ryssens}\ \emph {et~al.}(2023)\citenamefont
  {Ryssens}, \citenamefont {Giacalone}, \citenamefont {Schenke},\ and\
  \citenamefont {Shen}}]{ryssens2023}%
  \BibitemOpen
  \bibfield  {author} {\bibinfo {author} {\bibfnamefont {W.}~\bibnamefont
  {Ryssens}}, \bibinfo {author} {\bibfnamefont {G.}~\bibnamefont {Giacalone}},
  \bibinfo {author} {\bibfnamefont {B.}~\bibnamefont {Schenke}},\ and\ \bibinfo
  {author} {\bibfnamefont {C.}~\bibnamefont {Shen}},\ }\href
  {https://doi.org/10.1103/PhysRevLett.130.212302} {\bibfield  {journal}
  {\bibinfo  {journal} {Phys. Rev. Lett.}\ }\textbf {\bibinfo {volume} {130}},\
  \bibinfo {pages} {212302} (\bibinfo {year} {2023})}\BibitemShut {NoStop}%
\bibitem [{\citenamefont {Xu}\ \emph {et~al.}(2024)\citenamefont {Xu},
  \citenamefont {Zhao},\ and\ \citenamefont {Wang}}]{xu2024}%
  \BibitemOpen
  \bibfield  {author} {\bibinfo {author} {\bibfnamefont {H.-j.}\ \bibnamefont
  {Xu}}, \bibinfo {author} {\bibfnamefont {J.}~\bibnamefont {Zhao}},\ and\
  \bibinfo {author} {\bibfnamefont {F.}~\bibnamefont {Wang}},\ }\href
  {https://doi.org/10.1103/PhysRevLett.132.262301} {\bibfield  {journal}
  {\bibinfo  {journal} {Phys. Rev. Lett.}\ }\textbf {\bibinfo {volume} {132}},\
  \bibinfo {pages} {262301} (\bibinfo {year} {2024})}\BibitemShut {NoStop}%
\bibitem [{\citenamefont {Schenke}(2024)}]{schenke2024}%
  \BibitemOpen
  \bibfield  {author} {\bibinfo {author} {\bibfnamefont {B.}~\bibnamefont
  {Schenke}},\ }\href {https://doi.org/10.1007/s41365-024-01509-y} {\bibfield
  {journal} {\bibinfo  {journal} {Nuclear Science and Techniques}\ }\textbf
  {\bibinfo {volume} {35}},\ \bibinfo {pages} {115} (\bibinfo {year}
  {2024})}\BibitemShut {NoStop}%
\bibitem [{\citenamefont {Iachello}\ and\ \citenamefont {Arima}(1987)}]{IBM}%
  \BibitemOpen
  \bibfield  {author} {\bibinfo {author} {\bibfnamefont {F.}~\bibnamefont
  {Iachello}}\ and\ \bibinfo {author} {\bibfnamefont {A.}~\bibnamefont
  {Arima}},\ }\href@noop {} {\emph {\bibinfo {title} {The interacting boson
  model}}}\ (\bibinfo  {publisher} {Cambridge University Press, Cambridge},\
  \bibinfo {year} {1987})\BibitemShut {NoStop}%
\bibitem [{\citenamefont {Otsuka}\ \emph
  {et~al.}(1978{\natexlab{a}})\citenamefont {Otsuka}, \citenamefont {Arima},
  \citenamefont {Iachello},\ and\ \citenamefont {Talmi}}]{OAIT}%
  \BibitemOpen
  \bibfield  {author} {\bibinfo {author} {\bibfnamefont {T.}~\bibnamefont
  {Otsuka}}, \bibinfo {author} {\bibfnamefont {A.}~\bibnamefont {Arima}},
  \bibinfo {author} {\bibfnamefont {F.}~\bibnamefont {Iachello}},\ and\
  \bibinfo {author} {\bibfnamefont {I.}~\bibnamefont {Talmi}},\ }\href
  {https://doi.org/10.1016/0370-2693(78)90260-5} {\bibfield  {journal}
  {\bibinfo  {journal} {Phys. Lett. B}\ }\textbf {\bibinfo {volume} {76}},\
  \bibinfo {pages} {139 } (\bibinfo {year} {1978}{\natexlab{a}})}\BibitemShut
  {NoStop}%
\bibitem [{\citenamefont {Otsuka}\ \emph
  {et~al.}(1978{\natexlab{b}})\citenamefont {Otsuka}, \citenamefont {Arima},\
  and\ \citenamefont {Iachello}}]{OAI}%
  \BibitemOpen
  \bibfield  {author} {\bibinfo {author} {\bibfnamefont {T.}~\bibnamefont
  {Otsuka}}, \bibinfo {author} {\bibfnamefont {A.}~\bibnamefont {Arima}},\ and\
  \bibinfo {author} {\bibfnamefont {F.}~\bibnamefont {Iachello}},\ }\href
  {https://doi.org/10.1016/0375-9474(78)90532-8} {\bibfield  {journal}
  {\bibinfo  {journal} {Nucl. Phys. A}\ }\textbf {\bibinfo {volume} {309}},\
  \bibinfo {pages} {1} (\bibinfo {year} {1978}{\natexlab{b}})}\BibitemShut
  {NoStop}%
\bibitem [{\citenamefont {Casten}\ and\ \citenamefont
  {Warner}(1988)}]{casten1988}%
  \BibitemOpen
  \bibfield  {author} {\bibinfo {author} {\bibfnamefont {R.~F.}\ \bibnamefont
  {Casten}}\ and\ \bibinfo {author} {\bibfnamefont {D.~D.}\ \bibnamefont
  {Warner}},\ }\href {https://doi.org/10.1103/RevModPhys.60.389} {\bibfield
  {journal} {\bibinfo  {journal} {Rev. Mod. Phys.}\ }\textbf {\bibinfo {volume}
  {60}},\ \bibinfo {pages} {389} (\bibinfo {year} {1988})}\BibitemShut
  {NoStop}%
\bibitem [{\citenamefont {Otsuka}(1981)}]{otsuka1981}%
  \BibitemOpen
  \bibfield  {author} {\bibinfo {author} {\bibfnamefont {T.}~\bibnamefont
  {Otsuka}},\ }\href
  {https://doi.org/https://doi.org/10.1016/0375-9474(81)90685-0} {\bibfield
  {journal} {\bibinfo  {journal} {Nucl. Phys. A}\ }\textbf {\bibinfo {volume}
  {368}},\ \bibinfo {pages} {244} (\bibinfo {year} {1981})}\BibitemShut
  {NoStop}%
\bibitem [{\citenamefont {Otsuka}\ \emph {et~al.}(1982)\citenamefont {Otsuka},
  \citenamefont {Arima},\ and\ \citenamefont {Yoshinaga}}]{otsuka1982}%
  \BibitemOpen
  \bibfield  {author} {\bibinfo {author} {\bibfnamefont {T.}~\bibnamefont
  {Otsuka}}, \bibinfo {author} {\bibfnamefont {A.}~\bibnamefont {Arima}},\ and\
  \bibinfo {author} {\bibfnamefont {N.}~\bibnamefont {Yoshinaga}},\ }\href
  {https://doi.org/10.1103/PhysRevLett.48.387} {\bibfield  {journal} {\bibinfo
  {journal} {Phys. Rev. Lett.}\ }\textbf {\bibinfo {volume} {48}},\ \bibinfo
  {pages} {387} (\bibinfo {year} {1982})}\BibitemShut {NoStop}%
\bibitem [{\citenamefont {Otsuka}\ and\ \citenamefont
  {Ginocchio}(1985)}]{otsuka1985}%
  \BibitemOpen
  \bibfield  {author} {\bibinfo {author} {\bibfnamefont {T.}~\bibnamefont
  {Otsuka}}\ and\ \bibinfo {author} {\bibfnamefont {J.~N.}\ \bibnamefont
  {Ginocchio}},\ }\href {https://doi.org/10.1103/PhysRevLett.55.276} {\bibfield
   {journal} {\bibinfo  {journal} {Phys. Rev. Lett.}\ }\textbf {\bibinfo
  {volume} {55}},\ \bibinfo {pages} {276} (\bibinfo {year} {1985})}\BibitemShut
  {NoStop}%
\bibitem [{\citenamefont {Otsuka}\ and\ \citenamefont
  {Sugita}(1988)}]{otsuka1988}%
  \BibitemOpen
  \bibfield  {author} {\bibinfo {author} {\bibfnamefont {T.}~\bibnamefont
  {Otsuka}}\ and\ \bibinfo {author} {\bibfnamefont {M.}~\bibnamefont
  {Sugita}},\ }\href {https://doi.org/10.1016/0370-2693(88)90920-3} {\bibfield
  {journal} {\bibinfo  {journal} {Phys. Lett. B}\ }\textbf {\bibinfo {volume}
  {209}},\ \bibinfo {pages} {140 } (\bibinfo {year} {1988})}\BibitemShut
  {NoStop}%
\bibitem [{\citenamefont {Devi}\ and\ \citenamefont
  {Kota}(1990)}]{devi-kota1990}%
  \BibitemOpen
  \bibfield  {author} {\bibinfo {author} {\bibfnamefont {Y.~D.}\ \bibnamefont
  {Devi}}\ and\ \bibinfo {author} {\bibfnamefont {V.~K.~B.}\ \bibnamefont
  {Kota}},\ }\href {https://doi.org/10.1007/BF01283933} {\bibfield  {journal}
  {\bibinfo  {journal} {Z. Phys. A}\ }\textbf {\bibinfo {volume} {337}},\
  \bibinfo {pages} {15} (\bibinfo {year} {1990})}\BibitemShut {NoStop}%
\bibitem [{\citenamefont {Kuyucak}(1994)}]{kuyucak1994}%
  \BibitemOpen
  \bibfield  {author} {\bibinfo {author} {\bibfnamefont {S.}~\bibnamefont
  {Kuyucak}},\ }\href
  {https://doi.org/https://doi.org/10.1016/0375-9474(94)90282-8} {\bibfield
  {journal} {\bibinfo  {journal} {Nucl. Phys. A}\ }\textbf {\bibinfo {volume}
  {570}},\ \bibinfo {pages} {187} (\bibinfo {year} {1994})}\BibitemShut
  {NoStop}%
\bibitem [{\citenamefont {{Van Isacker}}\ \emph {et~al.}(2010)\citenamefont
  {{Van Isacker}}, \citenamefont {Bouldjedri},\ and\ \citenamefont
  {Zerguine}}]{vanisacker2010}%
  \BibitemOpen
  \bibfield  {author} {\bibinfo {author} {\bibfnamefont {P.}~\bibnamefont {{Van
  Isacker}}}, \bibinfo {author} {\bibfnamefont {A.}~\bibnamefont
  {Bouldjedri}},\ and\ \bibinfo {author} {\bibfnamefont {S.}~\bibnamefont
  {Zerguine}},\ }\href
  {https://doi.org/https://doi.org/10.1016/j.nuclphysa.2010.01.247} {\bibfield
  {journal} {\bibinfo  {journal} {Nucl. Phys. A}\ }\textbf {\bibinfo {volume}
  {836}},\ \bibinfo {pages} {225} (\bibinfo {year} {2010})}\BibitemShut
  {NoStop}%
\bibitem [{\citenamefont {Nomura}\ \emph {et~al.}(2008)\citenamefont {Nomura},
  \citenamefont {Shimizu},\ and\ \citenamefont {Otsuka}}]{nomura2008}%
  \BibitemOpen
  \bibfield  {author} {\bibinfo {author} {\bibfnamefont {K.}~\bibnamefont
  {Nomura}}, \bibinfo {author} {\bibfnamefont {N.}~\bibnamefont {Shimizu}},\
  and\ \bibinfo {author} {\bibfnamefont {T.}~\bibnamefont {Otsuka}},\ }\href
  {https://doi.org/10.1103/PhysRevLett.101.142501} {\bibfield  {journal}
  {\bibinfo  {journal} {Phys. Rev. Lett.}\ }\textbf {\bibinfo {volume} {101}},\
  \bibinfo {pages} {142501} (\bibinfo {year} {2008})}\BibitemShut {NoStop}%
\bibitem [{\citenamefont {Nomura}(2025)}]{nomura2025rev}%
  \BibitemOpen
  \bibfield  {author} {\bibinfo {author} {\bibfnamefont {K.}~\bibnamefont
  {Nomura}},\ }\href {https://doi.org/10.1140/epja/s10050-025-01604-7}
  {\bibfield  {journal} {\bibinfo  {journal} {Eur. Phys. J. A}\ }\textbf
  {\bibinfo {volume} {61}},\ \bibinfo {pages} {139} (\bibinfo {year}
  {2025})}\BibitemShut {NoStop}%
\bibitem [{\citenamefont {Bender}\ \emph {et~al.}(2003)\citenamefont {Bender},
  \citenamefont {Heenen},\ and\ \citenamefont {Reinhard}}]{bender2003}%
  \BibitemOpen
  \bibfield  {author} {\bibinfo {author} {\bibfnamefont {M.}~\bibnamefont
  {Bender}}, \bibinfo {author} {\bibfnamefont {P.-H.}\ \bibnamefont {Heenen}},\
  and\ \bibinfo {author} {\bibfnamefont {P.-G.}\ \bibnamefont {Reinhard}},\
  }\href {https://doi.org/10.1103/RevModPhys.75.121} {\bibfield  {journal}
  {\bibinfo  {journal} {Rev. Mod. Phys.}\ }\textbf {\bibinfo {volume} {75}},\
  \bibinfo {pages} {121} (\bibinfo {year} {2003})}\BibitemShut {NoStop}%
\bibitem [{\citenamefont {Vretenar}\ \emph {et~al.}(2005)\citenamefont
  {Vretenar}, \citenamefont {Afanasjev}, \citenamefont {Lalazissis},\ and\
  \citenamefont {Ring}}]{vretenar2005}%
  \BibitemOpen
  \bibfield  {author} {\bibinfo {author} {\bibfnamefont {D.}~\bibnamefont
  {Vretenar}}, \bibinfo {author} {\bibfnamefont {A.~V.}\ \bibnamefont
  {Afanasjev}}, \bibinfo {author} {\bibfnamefont {G.~A.}\ \bibnamefont
  {Lalazissis}},\ and\ \bibinfo {author} {\bibfnamefont {P.}~\bibnamefont
  {Ring}},\ }\href {https://doi.org/10.1016/j.physrep.2004.10.001} {\bibfield
  {journal} {\bibinfo  {journal} {Phys. Rep.}\ }\textbf {\bibinfo {volume}
  {409}},\ \bibinfo {pages} {101 } (\bibinfo {year} {2005})}\BibitemShut
  {NoStop}%
\bibitem [{\citenamefont {Robledo}\ \emph {et~al.}(2019)\citenamefont
  {Robledo}, \citenamefont {Rodríguez},\ and\ \citenamefont
  {Rodr{\'{i}}guez-Guzm{\'{a}}n}}]{robledo2019}%
  \BibitemOpen
  \bibfield  {author} {\bibinfo {author} {\bibfnamefont {L.~M.}\ \bibnamefont
  {Robledo}}, \bibinfo {author} {\bibfnamefont {T.~R.}\ \bibnamefont
  {Rodríguez}},\ and\ \bibinfo {author} {\bibfnamefont {R.~R.}\ \bibnamefont
  {Rodr{\'{i}}guez-Guzm{\'{a}}n}},\ }\href
  {http://stacks.iop.org/0954-3899/46/i=1/a=013001} {\bibfield  {journal}
  {\bibinfo  {journal} {J. Phys. G: Nucl. Part. Phys.}\ }\textbf {\bibinfo
  {volume} {46}},\ \bibinfo {pages} {013001} (\bibinfo {year}
  {2019})}\BibitemShut {NoStop}%
\bibitem [{\citenamefont {Nomura}\ \emph {et~al.}(2010)\citenamefont {Nomura},
  \citenamefont {Shimizu},\ and\ \citenamefont {Otsuka}}]{nomura2010}%
  \BibitemOpen
  \bibfield  {author} {\bibinfo {author} {\bibfnamefont {K.}~\bibnamefont
  {Nomura}}, \bibinfo {author} {\bibfnamefont {N.}~\bibnamefont {Shimizu}},\
  and\ \bibinfo {author} {\bibfnamefont {T.}~\bibnamefont {Otsuka}},\ }\href
  {https://doi.org/10.1103/PhysRevC.81.044307} {\bibfield  {journal} {\bibinfo
  {journal} {Phys. Rev. C}\ }\textbf {\bibinfo {volume} {81}},\ \bibinfo
  {pages} {044307} (\bibinfo {year} {2010})}\BibitemShut {NoStop}%
\bibitem [{\citenamefont {Nomura}\ \emph
  {et~al.}(2011{\natexlab{a}})\citenamefont {Nomura}, \citenamefont {Otsuka},
  \citenamefont {Rodr\'iguez-Guzm\'an}, \citenamefont {Robledo},\ and\
  \citenamefont {Sarriguren}}]{nomura2011pt}%
  \BibitemOpen
  \bibfield  {author} {\bibinfo {author} {\bibfnamefont {K.}~\bibnamefont
  {Nomura}}, \bibinfo {author} {\bibfnamefont {T.}~\bibnamefont {Otsuka}},
  \bibinfo {author} {\bibfnamefont {R.}~\bibnamefont {Rodr\'iguez-Guzm\'an}},
  \bibinfo {author} {\bibfnamefont {L.~M.}\ \bibnamefont {Robledo}},\ and\
  \bibinfo {author} {\bibfnamefont {P.}~\bibnamefont {Sarriguren}},\ }\href
  {https://doi.org/10.1103/PhysRevC.83.014309} {\bibfield  {journal} {\bibinfo
  {journal} {Phys. Rev. C}\ }\textbf {\bibinfo {volume} {83}},\ \bibinfo
  {pages} {014309} (\bibinfo {year} {2011}{\natexlab{a}})}\BibitemShut
  {NoStop}%
\bibitem [{\citenamefont {Nomura}\ \emph {et~al.}(2012)\citenamefont {Nomura},
  \citenamefont {Shimizu}, \citenamefont {Vretenar}, \citenamefont {Nik\ifmmode
  \check{s}\else \v{s}\fi{}i\ifmmode~\acute{c}\else \'{c}\fi{}},\ and\
  \citenamefont {Otsuka}}]{nomura2012tri}%
  \BibitemOpen
  \bibfield  {author} {\bibinfo {author} {\bibfnamefont {K.}~\bibnamefont
  {Nomura}}, \bibinfo {author} {\bibfnamefont {N.}~\bibnamefont {Shimizu}},
  \bibinfo {author} {\bibfnamefont {D.}~\bibnamefont {Vretenar}}, \bibinfo
  {author} {\bibfnamefont {T.}~\bibnamefont {Nik\ifmmode \check{s}\else
  \v{s}\fi{}i\ifmmode~\acute{c}\else \'{c}\fi{}}},\ and\ \bibinfo {author}
  {\bibfnamefont {T.}~\bibnamefont {Otsuka}},\ }\href
  {https://doi.org/10.1103/PhysRevLett.108.132501} {\bibfield  {journal}
  {\bibinfo  {journal} {Phys. Rev. Lett.}\ }\textbf {\bibinfo {volume} {108}},\
  \bibinfo {pages} {132501} (\bibinfo {year} {2012})}\BibitemShut {NoStop}%
\bibitem [{\citenamefont {Nomura}\ \emph {et~al.}(2013)\citenamefont {Nomura},
  \citenamefont {Vretenar},\ and\ \citenamefont {Lu}}]{nomura2013oct}%
  \BibitemOpen
  \bibfield  {author} {\bibinfo {author} {\bibfnamefont {K.}~\bibnamefont
  {Nomura}}, \bibinfo {author} {\bibfnamefont {D.}~\bibnamefont {Vretenar}},\
  and\ \bibinfo {author} {\bibfnamefont {B.-N.}\ \bibnamefont {Lu}},\ }\href
  {https://doi.org/10.1103/PhysRevC.88.021303} {\bibfield  {journal} {\bibinfo
  {journal} {Phys. Rev. C}\ }\textbf {\bibinfo {volume} {88}},\ \bibinfo
  {pages} {021303} (\bibinfo {year} {2013})}\BibitemShut {NoStop}%
\bibitem [{\citenamefont {Nomura}\ \emph {et~al.}(2014)\citenamefont {Nomura},
  \citenamefont {Vretenar}, \citenamefont {Nik\ifmmode \check{s}\else
  \v{s}\fi{}i\ifmmode~\acute{c}\else \'{c}\fi{}},\ and\ \citenamefont
  {Lu}}]{nomura2014}%
  \BibitemOpen
  \bibfield  {author} {\bibinfo {author} {\bibfnamefont {K.}~\bibnamefont
  {Nomura}}, \bibinfo {author} {\bibfnamefont {D.}~\bibnamefont {Vretenar}},
  \bibinfo {author} {\bibfnamefont {T.}~\bibnamefont {Nik\ifmmode
  \check{s}\else \v{s}\fi{}i\ifmmode~\acute{c}\else \'{c}\fi{}}},\ and\
  \bibinfo {author} {\bibfnamefont {B.-N.}\ \bibnamefont {Lu}},\ }\href
  {https://doi.org/10.1103/PhysRevC.89.024312} {\bibfield  {journal} {\bibinfo
  {journal} {Phys. Rev. C}\ }\textbf {\bibinfo {volume} {89}},\ \bibinfo
  {pages} {024312} (\bibinfo {year} {2014})}\BibitemShut {NoStop}%
\bibitem [{\citenamefont {Nomura}(2023)}]{nomura2023oct}%
  \BibitemOpen
  \bibfield  {author} {\bibinfo {author} {\bibfnamefont {K.}~\bibnamefont
  {Nomura}},\ }\href {https://doi.org/10.1142/S0218301323400013} {\bibfield
  {journal} {\bibinfo  {journal} {Int. J. Mod. Phys. E}\ }\textbf {\bibinfo
  {volume} {32}},\ \bibinfo {pages} {2340001} (\bibinfo {year}
  {2023})}\BibitemShut {NoStop}%
\bibitem [{\citenamefont {Lotina}\ and\ \citenamefont
  {Nomura}(2024{\natexlab{a}})}]{lotina2024hex-1}%
  \BibitemOpen
  \bibfield  {author} {\bibinfo {author} {\bibfnamefont {L.}~\bibnamefont
  {Lotina}}\ and\ \bibinfo {author} {\bibfnamefont {K.}~\bibnamefont
  {Nomura}},\ }\href {https://doi.org/10.1103/PhysRevC.109.034304} {\bibfield
  {journal} {\bibinfo  {journal} {Phys. Rev. C}\ }\textbf {\bibinfo {volume}
  {109}},\ \bibinfo {pages} {034304} (\bibinfo {year}
  {2024}{\natexlab{a}})}\BibitemShut {NoStop}%
\bibitem [{\citenamefont {Lotina}\ and\ \citenamefont
  {Nomura}(2024{\natexlab{b}})}]{lotina2024hex-2}%
  \BibitemOpen
  \bibfield  {author} {\bibinfo {author} {\bibfnamefont {L.}~\bibnamefont
  {Lotina}}\ and\ \bibinfo {author} {\bibfnamefont {K.}~\bibnamefont
  {Nomura}},\ }\href {https://doi.org/10.1103/PhysRevC.109.044324} {\bibfield
  {journal} {\bibinfo  {journal} {Phys. Rev. C}\ }\textbf {\bibinfo {volume}
  {109}},\ \bibinfo {pages} {044324} (\bibinfo {year}
  {2024}{\natexlab{b}})}\BibitemShut {NoStop}%
\bibitem [{\citenamefont {Lotina}\ \emph {et~al.}(2025)\citenamefont {Lotina},
  \citenamefont {Nomura}, \citenamefont {Rodr\'{\i}guez-Guzm\'an},\ and\
  \citenamefont {Robledo}}]{lotina2025}%
  \BibitemOpen
  \bibfield  {author} {\bibinfo {author} {\bibfnamefont {L.}~\bibnamefont
  {Lotina}}, \bibinfo {author} {\bibfnamefont {K.}~\bibnamefont {Nomura}},
  \bibinfo {author} {\bibfnamefont {R.}~\bibnamefont
  {Rodr\'{\i}guez-Guzm\'an}},\ and\ \bibinfo {author} {\bibfnamefont {L.~M.}\
  \bibnamefont {Robledo}},\ }\href
  {https://doi.org/10.1103/PhysRevC.111.024301} {\bibfield  {journal} {\bibinfo
   {journal} {Phys. Rev. C}\ }\textbf {\bibinfo {volume} {111}},\ \bibinfo
  {pages} {024301} (\bibinfo {year} {2025})}\BibitemShut {NoStop}%
\bibitem [{\citenamefont {Gogny}(1975)}]{GOGNY1975}%
  \BibitemOpen
  \bibfield  {author} {\bibinfo {author} {\bibfnamefont {D.}~\bibnamefont
  {Gogny}},\ }\href
  {https://doi.org/https://doi.org/10.1016/0375-9474(75)90407-8} {\bibfield
  {journal} {\bibinfo  {journal} {Nuclear Physics A}\ }\textbf {\bibinfo
  {volume} {237}},\ \bibinfo {pages} {399} (\bibinfo {year}
  {1975})}\BibitemShut {NoStop}%
\bibitem [{\citenamefont {Berger}\ \emph {et~al.}(1984)\citenamefont {Berger},
  \citenamefont {Girod},\ and\ \citenamefont {Gogny}}]{BERGER1984}%
  \BibitemOpen
  \bibfield  {author} {\bibinfo {author} {\bibfnamefont {J.}~\bibnamefont
  {Berger}}, \bibinfo {author} {\bibfnamefont {M.}~\bibnamefont {Girod}},\ and\
  \bibinfo {author} {\bibfnamefont {D.}~\bibnamefont {Gogny}},\ }\href
  {https://doi.org/https://doi.org/10.1016/0375-9474(84)90240-9} {\bibfield
  {journal} {\bibinfo  {journal} {Nuclear Physics A}\ }\textbf {\bibinfo
  {volume} {428}},\ \bibinfo {pages} {23} (\bibinfo {year} {1984})}\BibitemShut
  {NoStop}%
\bibitem [{\citenamefont {Hilaire}\ and\ \citenamefont
  {Girod}(2007)}]{hilaire2007}%
  \BibitemOpen
  \bibfield  {author} {\bibinfo {author} {\bibfnamefont {S.}~\bibnamefont
  {Hilaire}}\ and\ \bibinfo {author} {\bibfnamefont {M.}~\bibnamefont
  {Girod}},\ }\href
  {https://doi.org/https://doi.org/10.1140/epja/i2007-10450-2} {\bibfield
  {journal} {\bibinfo  {journal} {Eur. Phys. J. A}\ }\textbf {\bibinfo {volume}
  {33}},\ \bibinfo {pages} {237} (\bibinfo {year} {2007})}\BibitemShut
  {NoStop}%
\bibitem [{\citenamefont {Nomura}\ \emph
  {et~al.}(2011{\natexlab{b}})\citenamefont {Nomura}, \citenamefont {Otsuka},
  \citenamefont {Rodr\'{\i}guez-Guzm\'an}, \citenamefont {Robledo},
  \citenamefont {Sarriguren}, \citenamefont {Regan}, \citenamefont
  {Stevenson},\ and\ \citenamefont {Podoly\'ak}}]{nomura2011Os}%
  \BibitemOpen
  \bibfield  {author} {\bibinfo {author} {\bibfnamefont {K.}~\bibnamefont
  {Nomura}}, \bibinfo {author} {\bibfnamefont {T.}~\bibnamefont {Otsuka}},
  \bibinfo {author} {\bibfnamefont {R.}~\bibnamefont
  {Rodr\'{\i}guez-Guzm\'an}}, \bibinfo {author} {\bibfnamefont {L.~M.}\
  \bibnamefont {Robledo}}, \bibinfo {author} {\bibfnamefont {P.}~\bibnamefont
  {Sarriguren}}, \bibinfo {author} {\bibfnamefont {P.~H.}\ \bibnamefont
  {Regan}}, \bibinfo {author} {\bibfnamefont {P.~D.}\ \bibnamefont
  {Stevenson}},\ and\ \bibinfo {author} {\bibfnamefont {Z.}~\bibnamefont
  {Podoly\'ak}},\ }\href {https://doi.org/10.1103/PhysRevC.83.054303}
  {\bibfield  {journal} {\bibinfo  {journal} {Phys. Rev. C}\ }\textbf {\bibinfo
  {volume} {83}},\ \bibinfo {pages} {054303} (\bibinfo {year}
  {2011}{\natexlab{b}})}\BibitemShut {NoStop}%
\bibitem [{\citenamefont {Nomura}\ \emph
  {et~al.}(2011{\natexlab{c}})\citenamefont {Nomura}, \citenamefont {Otsuka},
  \citenamefont {Rodr\'iguez-Guzm\'an}, \citenamefont {Robledo},\ and\
  \citenamefont {Sarriguren}}]{nomura2011sys}%
  \BibitemOpen
  \bibfield  {author} {\bibinfo {author} {\bibfnamefont {K.}~\bibnamefont
  {Nomura}}, \bibinfo {author} {\bibfnamefont {T.}~\bibnamefont {Otsuka}},
  \bibinfo {author} {\bibfnamefont {R.}~\bibnamefont {Rodr\'iguez-Guzm\'an}},
  \bibinfo {author} {\bibfnamefont {L.~M.}\ \bibnamefont {Robledo}},\ and\
  \bibinfo {author} {\bibfnamefont {P.}~\bibnamefont {Sarriguren}},\ }\href
  {https://doi.org/10.1103/PhysRevC.84.054316} {\bibfield  {journal} {\bibinfo
  {journal} {Phys. Rev. C}\ }\textbf {\bibinfo {volume} {84}},\ \bibinfo
  {pages} {054316} (\bibinfo {year} {2011}{\natexlab{c}})}\BibitemShut
  {NoStop}%
\bibitem [{\citenamefont {Nomura}\ \emph
  {et~al.}(2017{\natexlab{a}})\citenamefont {Nomura}, \citenamefont
  {Rodr\'{\i}guez-Guzm\'an},\ and\ \citenamefont {Robledo}}]{nomura2017ge}%
  \BibitemOpen
  \bibfield  {author} {\bibinfo {author} {\bibfnamefont {K.}~\bibnamefont
  {Nomura}}, \bibinfo {author} {\bibfnamefont {R.}~\bibnamefont
  {Rodr\'{\i}guez-Guzm\'an}},\ and\ \bibinfo {author} {\bibfnamefont {L.~M.}\
  \bibnamefont {Robledo}},\ }\href {https://doi.org/10.1103/PhysRevC.95.064310}
  {\bibfield  {journal} {\bibinfo  {journal} {Phys. Rev. C}\ }\textbf {\bibinfo
  {volume} {95}},\ \bibinfo {pages} {064310} (\bibinfo {year}
  {2017}{\natexlab{a}})}\BibitemShut {NoStop}%
\bibitem [{\citenamefont {Nomura}\ \emph
  {et~al.}(2017{\natexlab{b}})\citenamefont {Nomura}, \citenamefont
  {Rodr\'{\i}guez-Guzm\'an}, \citenamefont {Humadi}, \citenamefont {Robledo},\
  and\ \citenamefont {Abusara}}]{nomura2017kr}%
  \BibitemOpen
  \bibfield  {author} {\bibinfo {author} {\bibfnamefont {K.}~\bibnamefont
  {Nomura}}, \bibinfo {author} {\bibfnamefont {R.}~\bibnamefont
  {Rodr\'{\i}guez-Guzm\'an}}, \bibinfo {author} {\bibfnamefont {Y.~M.}\
  \bibnamefont {Humadi}}, \bibinfo {author} {\bibfnamefont {L.~M.}\
  \bibnamefont {Robledo}},\ and\ \bibinfo {author} {\bibfnamefont
  {H.}~\bibnamefont {Abusara}},\ }\href
  {https://doi.org/10.1103/PhysRevC.96.034310} {\bibfield  {journal} {\bibinfo
  {journal} {Phys. Rev. C}\ }\textbf {\bibinfo {volume} {96}},\ \bibinfo
  {pages} {034310} (\bibinfo {year} {2017}{\natexlab{b}})}\BibitemShut
  {NoStop}%
\bibitem [{\citenamefont {Nomura}\ \emph {et~al.}(2015)\citenamefont {Nomura},
  \citenamefont {Rodr\'{\i}guez-Guzm\'an},\ and\ \citenamefont
  {Robledo}}]{nomura2015}%
  \BibitemOpen
  \bibfield  {author} {\bibinfo {author} {\bibfnamefont {K.}~\bibnamefont
  {Nomura}}, \bibinfo {author} {\bibfnamefont {R.}~\bibnamefont
  {Rodr\'{\i}guez-Guzm\'an}},\ and\ \bibinfo {author} {\bibfnamefont {L.~M.}\
  \bibnamefont {Robledo}},\ }\href {https://doi.org/10.1103/PhysRevC.92.014312}
  {\bibfield  {journal} {\bibinfo  {journal} {Phys. Rev. C}\ }\textbf {\bibinfo
  {volume} {92}},\ \bibinfo {pages} {014312} (\bibinfo {year}
  {2015})}\BibitemShut {NoStop}%
\bibitem [{\citenamefont {Nomura}\ \emph {et~al.}(2020)\citenamefont {Nomura},
  \citenamefont {Rodr\'{\i}guez-Guzm\'an}, \citenamefont {Humadi},
  \citenamefont {Robledo},\ and\ \citenamefont
  {Garc\'{\i}a-Ramos}}]{nomura2020oct}%
  \BibitemOpen
  \bibfield  {author} {\bibinfo {author} {\bibfnamefont {K.}~\bibnamefont
  {Nomura}}, \bibinfo {author} {\bibfnamefont {R.}~\bibnamefont
  {Rodr\'{\i}guez-Guzm\'an}}, \bibinfo {author} {\bibfnamefont {Y.~M.}\
  \bibnamefont {Humadi}}, \bibinfo {author} {\bibfnamefont {L.~M.}\
  \bibnamefont {Robledo}},\ and\ \bibinfo {author} {\bibfnamefont {J.~E.}\
  \bibnamefont {Garc\'{\i}a-Ramos}},\ }\href
  {https://doi.org/10.1103/PhysRevC.102.064326} {\bibfield  {journal} {\bibinfo
   {journal} {Phys. Rev. C}\ }\textbf {\bibinfo {volume} {102}},\ \bibinfo
  {pages} {064326} (\bibinfo {year} {2020})}\BibitemShut {NoStop}%
\bibitem [{\citenamefont {Nomura}\ \emph
  {et~al.}(2021{\natexlab{a}})\citenamefont {Nomura}, \citenamefont
  {Rodr\'{\i}guez-Guzm\'an}, \citenamefont {Robledo},\ and\ \citenamefont
  {Garc\'{\i}a-Ramos}}]{nomura2021oct-u}%
  \BibitemOpen
  \bibfield  {author} {\bibinfo {author} {\bibfnamefont {K.}~\bibnamefont
  {Nomura}}, \bibinfo {author} {\bibfnamefont {R.}~\bibnamefont
  {Rodr\'{\i}guez-Guzm\'an}}, \bibinfo {author} {\bibfnamefont
  {L.}~\bibnamefont {Robledo}},\ and\ \bibinfo {author} {\bibfnamefont
  {J.}~\bibnamefont {Garc\'{\i}a-Ramos}},\ }\href
  {https://doi.org/10.1103/PhysRevC.103.044311} {\bibfield  {journal} {\bibinfo
   {journal} {Phys. Rev. C}\ }\textbf {\bibinfo {volume} {103}},\ \bibinfo
  {pages} {044311} (\bibinfo {year} {2021}{\natexlab{a}})}\BibitemShut
  {NoStop}%
\bibitem [{\citenamefont {Nomura}\ \emph
  {et~al.}(2021{\natexlab{b}})\citenamefont {Nomura}, \citenamefont
  {Rodr\'{\i}guez-Guzm\'an}, \citenamefont {Robledo}, \citenamefont
  {Garc\'{\i}a-Ramos},\ and\ \citenamefont {Hern\'andez}}]{nomura2021oct-ba}%
  \BibitemOpen
  \bibfield  {author} {\bibinfo {author} {\bibfnamefont {K.}~\bibnamefont
  {Nomura}}, \bibinfo {author} {\bibfnamefont {R.}~\bibnamefont
  {Rodr\'{\i}guez-Guzm\'an}}, \bibinfo {author} {\bibfnamefont {L.~M.}\
  \bibnamefont {Robledo}}, \bibinfo {author} {\bibfnamefont {J.~E.}\
  \bibnamefont {Garc\'{\i}a-Ramos}},\ and\ \bibinfo {author} {\bibfnamefont
  {N.~C.}\ \bibnamefont {Hern\'andez}},\ }\href
  {https://doi.org/10.1103/PhysRevC.104.044324} {\bibfield  {journal} {\bibinfo
   {journal} {Phys. Rev. C}\ }\textbf {\bibinfo {volume} {104}},\ \bibinfo
  {pages} {044324} (\bibinfo {year} {2021}{\natexlab{b}})}\BibitemShut
  {NoStop}%
\bibitem [{\citenamefont {Nomura}\ \emph
  {et~al.}(2021{\natexlab{c}})\citenamefont {Nomura}, \citenamefont
  {Rodr\'{\i}guez-Guzm\'an},\ and\ \citenamefont {Robledo}}]{nomura2021oct-zn}%
  \BibitemOpen
  \bibfield  {author} {\bibinfo {author} {\bibfnamefont {K.}~\bibnamefont
  {Nomura}}, \bibinfo {author} {\bibfnamefont {R.}~\bibnamefont
  {Rodr\'{\i}guez-Guzm\'an}},\ and\ \bibinfo {author} {\bibfnamefont {L.~M.}\
  \bibnamefont {Robledo}},\ }\href
  {https://doi.org/10.1103/PhysRevC.104.054320} {\bibfield  {journal} {\bibinfo
   {journal} {Phys. Rev. C}\ }\textbf {\bibinfo {volume} {104}},\ \bibinfo
  {pages} {054320} (\bibinfo {year} {2021}{\natexlab{c}})}\BibitemShut
  {NoStop}%
\bibitem [{\citenamefont {Kumar}\ and\ \citenamefont
  {Robledo}(2023)}]{kumar-robledo2023}%
  \BibitemOpen
  \bibfield  {author} {\bibinfo {author} {\bibfnamefont {C.~V.~N.}\
  \bibnamefont {Kumar}}\ and\ \bibinfo {author} {\bibfnamefont {L.~M.}\
  \bibnamefont {Robledo}},\ }\href
  {https://doi.org/10.1103/PhysRevC.108.034312} {\bibfield  {journal} {\bibinfo
   {journal} {Phys. Rev. C}\ }\textbf {\bibinfo {volume} {108}},\ \bibinfo
  {pages} {034312} (\bibinfo {year} {2023})}\BibitemShut {NoStop}%
\bibitem [{\citenamefont {Rodr\'{\i}guez-Guzm\'an}\ and\ \citenamefont
  {Robledo}(2025)}]{guzman2025}%
  \BibitemOpen
  \bibfield  {author} {\bibinfo {author} {\bibfnamefont {R.}~\bibnamefont
  {Rodr\'{\i}guez-Guzm\'an}}\ and\ \bibinfo {author} {\bibfnamefont {L.~M.}\
  \bibnamefont {Robledo}},\ }\href
  {https://doi.org/10.1103/PhysRevC.111.024304} {\bibfield  {journal} {\bibinfo
   {journal} {Phys. Rev. C}\ }\textbf {\bibinfo {volume} {111}},\ \bibinfo
  {pages} {024304} (\bibinfo {year} {2025})}\BibitemShut {NoStop}%
\bibitem [{\citenamefont {Rodr{\'i}guez-Guzm{\'a}n}\ and\ \citenamefont
  {Robledo}(2025)}]{guzman2025rare_earth}%
  \BibitemOpen
  \bibfield  {author} {\bibinfo {author} {\bibfnamefont {R.}~\bibnamefont
  {Rodr{\'i}guez-Guzm{\'a}n}}\ and\ \bibinfo {author} {\bibfnamefont {L.~M.}\
  \bibnamefont {Robledo}},\ }\href
  {https://doi.org/10.1140/epja/s10050-025-01638-x} {\bibfield  {journal}
  {\bibinfo  {journal} {The European Physical Journal A}\ }\textbf {\bibinfo
  {volume} {61}},\ \bibinfo {pages} {166} (\bibinfo {year} {2025})}\BibitemShut
  {NoStop}%
\bibitem [{\citenamefont {Ring}\ and\ \citenamefont {Schuck}(1980)}]{RS}%
  \BibitemOpen
  \bibfield  {author} {\bibinfo {author} {\bibfnamefont {P.}~\bibnamefont
  {Ring}}\ and\ \bibinfo {author} {\bibfnamefont {P.}~\bibnamefont {Schuck}},\
  }\href@noop {} {\emph {\bibinfo {title} {The Nuclear Many-Body Problem}}}\
  (\bibinfo  {publisher} {Springer-Verlag, Berlin},\ \bibinfo {year}
  {1980})\BibitemShut {NoStop}%
\bibitem [{\citenamefont {Nik\ifmmode \check{s}\else
  \v{s}\fi{}i\ifmmode~\acute{c}\else \'{c}\fi{}}\ \emph
  {et~al.}(2011)\citenamefont {Nik\ifmmode \check{s}\else
  \v{s}\fi{}i\ifmmode~\acute{c}\else \'{c}\fi{}}, \citenamefont {Vretenar},\
  and\ \citenamefont {Ring}}]{niksic2011}%
  \BibitemOpen
  \bibfield  {author} {\bibinfo {author} {\bibfnamefont {T.}~\bibnamefont
  {Nik\ifmmode \check{s}\else \v{s}\fi{}i\ifmmode~\acute{c}\else \'{c}\fi{}}},
  \bibinfo {author} {\bibfnamefont {D.}~\bibnamefont {Vretenar}},\ and\
  \bibinfo {author} {\bibfnamefont {P.}~\bibnamefont {Ring}},\ }\href
  {https://doi.org/10.1016/j.ppnp.2011.01.055} {\bibfield  {journal} {\bibinfo
  {journal} {Prog. Part. Nucl. Phys.}\ }\textbf {\bibinfo {volume} {66}},\
  \bibinfo {pages} {519} (\bibinfo {year} {2011})}\BibitemShut {NoStop}%
\bibitem [{\citenamefont {Butler}\ and\ \citenamefont
  {Nazarewicz}(1996)}]{butler1996}%
  \BibitemOpen
  \bibfield  {author} {\bibinfo {author} {\bibfnamefont {P.~A.}\ \bibnamefont
  {Butler}}\ and\ \bibinfo {author} {\bibfnamefont {W.}~\bibnamefont
  {Nazarewicz}},\ }\href {https://doi.org/10.1103/RevModPhys.68.349} {\bibfield
   {journal} {\bibinfo  {journal} {Rev. Mod. Phys.}\ }\textbf {\bibinfo
  {volume} {68}},\ \bibinfo {pages} {349} (\bibinfo {year} {1996})}\BibitemShut
  {NoStop}%
\bibitem [{\citenamefont {Butler}(2016)}]{butler2016}%
  \BibitemOpen
  \bibfield  {author} {\bibinfo {author} {\bibfnamefont {P.~A.}\ \bibnamefont
  {Butler}},\ }\href {https://doi.org/10.1088/0954-3899/43/7/073002} {\bibfield
   {journal} {\bibinfo  {journal} {J. Phys. G: Nucl. Part. Phys.}\ }\textbf
  {\bibinfo {volume} {43}},\ \bibinfo {pages} {073002} (\bibinfo {year}
  {2016})}\BibitemShut {NoStop}%
\bibitem [{\citenamefont {Spieker}\ \emph {et~al.}(2013)\citenamefont
  {Spieker}, \citenamefont {Bucurescu}, \citenamefont {Endres}, \citenamefont
  {Faestermann}, \citenamefont {Hertenberger}, \citenamefont {Pascu},
  \citenamefont {Skalacki}, \citenamefont {Weber}, \citenamefont {Wirth},
  \citenamefont {Zamfir},\ and\ \citenamefont {Zilges}}]{spieker2013}%
  \BibitemOpen
  \bibfield  {author} {\bibinfo {author} {\bibfnamefont {M.}~\bibnamefont
  {Spieker}}, \bibinfo {author} {\bibfnamefont {D.}~\bibnamefont {Bucurescu}},
  \bibinfo {author} {\bibfnamefont {J.}~\bibnamefont {Endres}}, \bibinfo
  {author} {\bibfnamefont {T.}~\bibnamefont {Faestermann}}, \bibinfo {author}
  {\bibfnamefont {R.}~\bibnamefont {Hertenberger}}, \bibinfo {author}
  {\bibfnamefont {S.}~\bibnamefont {Pascu}}, \bibinfo {author} {\bibfnamefont
  {S.}~\bibnamefont {Skalacki}}, \bibinfo {author} {\bibfnamefont
  {S.}~\bibnamefont {Weber}}, \bibinfo {author} {\bibfnamefont {H.-F.}\
  \bibnamefont {Wirth}}, \bibinfo {author} {\bibfnamefont {N.-V.}\ \bibnamefont
  {Zamfir}},\ and\ \bibinfo {author} {\bibfnamefont {A.}~\bibnamefont
  {Zilges}},\ }\href {https://doi.org/10.1103/PhysRevC.88.041303} {\bibfield
  {journal} {\bibinfo  {journal} {Phys. Rev. C}\ }\textbf {\bibinfo {volume}
  {88}},\ \bibinfo {pages} {041303} (\bibinfo {year} {2013})}\BibitemShut
  {NoStop}%
\bibitem [{\citenamefont {Spieker}\ \emph {et~al.}(2018)\citenamefont
  {Spieker}, \citenamefont {Pascu}, \citenamefont {Bucurescu}, \citenamefont
  {Shneidman}, \citenamefont {Faestermann}, \citenamefont {Hertenberger},
  \citenamefont {Wirth}, \citenamefont {Zamfir},\ and\ \citenamefont
  {Zilges}}]{spieker2018}%
  \BibitemOpen
  \bibfield  {author} {\bibinfo {author} {\bibfnamefont {M.}~\bibnamefont
  {Spieker}}, \bibinfo {author} {\bibfnamefont {S.}~\bibnamefont {Pascu}},
  \bibinfo {author} {\bibfnamefont {D.}~\bibnamefont {Bucurescu}}, \bibinfo
  {author} {\bibfnamefont {T.~M.}\ \bibnamefont {Shneidman}}, \bibinfo {author}
  {\bibfnamefont {T.}~\bibnamefont {Faestermann}}, \bibinfo {author}
  {\bibfnamefont {R.}~\bibnamefont {Hertenberger}}, \bibinfo {author}
  {\bibfnamefont {H.-F.}\ \bibnamefont {Wirth}}, \bibinfo {author}
  {\bibfnamefont {N.-V.}\ \bibnamefont {Zamfir}},\ and\ \bibinfo {author}
  {\bibfnamefont {A.}~\bibnamefont {Zilges}},\ }\href
  {https://doi.org/10.1103/PhysRevC.97.064319} {\bibfield  {journal} {\bibinfo
  {journal} {Phys. Rev. C}\ }\textbf {\bibinfo {volume} {97}},\ \bibinfo
  {pages} {064319} (\bibinfo {year} {2018})}\BibitemShut {NoStop}%
\bibitem [{\citenamefont {Kota}\ \emph {et~al.}(1987)\citenamefont {Kota},
  \citenamefont {der Jeugt}, \citenamefont {Meyer},\ and\ \citenamefont
  {Berghe}}]{kota1987}%
  \BibitemOpen
  \bibfield  {author} {\bibinfo {author} {\bibfnamefont {V.~K.~B.}\
  \bibnamefont {Kota}}, \bibinfo {author} {\bibfnamefont {J.~V.}\ \bibnamefont
  {der Jeugt}}, \bibinfo {author} {\bibfnamefont {H.~E.~D.}\ \bibnamefont
  {Meyer}},\ and\ \bibinfo {author} {\bibfnamefont {G.~V.}\ \bibnamefont
  {Berghe}},\ }\href {https://api.semanticscholar.org/CorpusID:120379709}
  {\bibfield  {journal} {\bibinfo  {journal} {Journal of Mathematical Physics}\
  }\textbf {\bibinfo {volume} {28}},\ \bibinfo {pages} {1644} (\bibinfo {year}
  {1987})}\BibitemShut {NoStop}%
\bibitem [{\citenamefont {Ginocchio}\ and\ \citenamefont
  {Kirson}(1980)}]{ginocchio1980}%
  \BibitemOpen
  \bibfield  {author} {\bibinfo {author} {\bibfnamefont {J.~N.}\ \bibnamefont
  {Ginocchio}}\ and\ \bibinfo {author} {\bibfnamefont {M.~W.}\ \bibnamefont
  {Kirson}},\ }\href {https://doi.org/10.1016/0375-9474(80)90387-5} {\bibfield
  {journal} {\bibinfo  {journal} {Nucl. Phys. A}\ }\textbf {\bibinfo {volume}
  {350}},\ \bibinfo {pages} {31} (\bibinfo {year} {1980})}\BibitemShut
  {NoStop}%
\bibitem [{\citenamefont {Dieperink}\ \emph {et~al.}(1980)\citenamefont
  {Dieperink}, \citenamefont {Scholten},\ and\ \citenamefont
  {Iachello}}]{dieperink1980}%
  \BibitemOpen
  \bibfield  {author} {\bibinfo {author} {\bibfnamefont {A.~E.~L.}\
  \bibnamefont {Dieperink}}, \bibinfo {author} {\bibfnamefont {O.}~\bibnamefont
  {Scholten}},\ and\ \bibinfo {author} {\bibfnamefont {F.}~\bibnamefont
  {Iachello}},\ }\href {https://doi.org/10.1103/PhysRevLett.44.1747} {\bibfield
   {journal} {\bibinfo  {journal} {Phys. Rev. Lett.}\ }\textbf {\bibinfo
  {volume} {44}},\ \bibinfo {pages} {1747} (\bibinfo {year}
  {1980})}\BibitemShut {NoStop}%
\bibitem [{\citenamefont {Heinze}()}]{arbmodel}%
  \BibitemOpen
  \bibfield  {author} {\bibinfo {author} {\bibfnamefont {S.}~\bibnamefont
  {Heinze}},\ }\href@noop {} {}\bibinfo {note} {{}computer program ARBMODEL,
  University of Cologne (2008)}\BibitemShut {NoStop}%
\bibitem [{\citenamefont {{Brookhaven National Nuclear Data Center}}()}]{data}%
  \BibitemOpen
  \bibfield  {author} {\bibinfo {author} {\bibnamefont {{Brookhaven National
  Nuclear Data Center}}},\ }\href@noop {} {}\bibinfo {howpublished}
  {{\url{http://www.nndc.bnl.gov}}}\BibitemShut {NoStop}%
\bibitem [{\citenamefont {Rodr\'{\i}guez-Guzm\'an}\ \emph
  {et~al.}(2010)\citenamefont {Rodr\'{\i}guez-Guzm\'an}, \citenamefont
  {Sarriguren}, \citenamefont {Robledo},\ and\ \citenamefont
  {Garc\'{\i}a-Ramos}}]{RAY-gamma-1}%
  \BibitemOpen
  \bibfield  {author} {\bibinfo {author} {\bibfnamefont {R.}~\bibnamefont
  {Rodr\'{\i}guez-Guzm\'an}}, \bibinfo {author} {\bibfnamefont
  {P.}~\bibnamefont {Sarriguren}}, \bibinfo {author} {\bibfnamefont {L.~M.}\
  \bibnamefont {Robledo}},\ and\ \bibinfo {author} {\bibfnamefont {J.~E.}\
  \bibnamefont {Garc\'{\i}a-Ramos}},\ }\href
  {https://doi.org/10.1103/PhysRevC.81.024310} {\bibfield  {journal} {\bibinfo
  {journal} {Phys. Rev. C}\ }\textbf {\bibinfo {volume} {81}},\ \bibinfo
  {pages} {024310} (\bibinfo {year} {2010})}\BibitemShut {NoStop}%
\bibitem [{\citenamefont {Robledo}\ \emph {et~al.}(2008)\citenamefont
  {Robledo}, \citenamefont {Rodr\'{\i}guez-Guzm\'an},\ and\ \citenamefont
  {Sarriguren}}]{RAY-gamma-2}%
  \BibitemOpen
  \bibfield  {author} {\bibinfo {author} {\bibfnamefont {L.~M.}\ \bibnamefont
  {Robledo}}, \bibinfo {author} {\bibfnamefont {R.~R.}\ \bibnamefont
  {Rodr\'{\i}guez-Guzm\'an}},\ and\ \bibinfo {author} {\bibfnamefont
  {P.}~\bibnamefont {Sarriguren}},\ }\href
  {https://doi.org/10.1103/PhysRevC.78.034314} {\bibfield  {journal} {\bibinfo
  {journal} {Phys. Rev. C}\ }\textbf {\bibinfo {volume} {78}},\ \bibinfo
  {pages} {034314} (\bibinfo {year} {2008})}\BibitemShut {NoStop}%
\bibitem [{\citenamefont {Lu}\ \emph {et~al.}(2014)\citenamefont {Lu},
  \citenamefont {Zhao}, \citenamefont {Zhao},\ and\ \citenamefont
  {Zhou}}]{lu2014}%
  \BibitemOpen
  \bibfield  {author} {\bibinfo {author} {\bibfnamefont {B.-N.}\ \bibnamefont
  {Lu}}, \bibinfo {author} {\bibfnamefont {J.}~\bibnamefont {Zhao}}, \bibinfo
  {author} {\bibfnamefont {E.-G.}\ \bibnamefont {Zhao}},\ and\ \bibinfo
  {author} {\bibfnamefont {S.-G.}\ \bibnamefont {Zhou}},\ }\href
  {https://doi.org/10.1103/PhysRevC.89.014323} {\bibfield  {journal} {\bibinfo
  {journal} {Phys. Rev. C}\ }\textbf {\bibinfo {volume} {89}},\ \bibinfo
  {pages} {014323} (\bibinfo {year} {2014})}\BibitemShut {NoStop}%
\end{thebibliography}%

\end{document}